\documentclass[usenatbib]{mn2e}
\usepackage{cite}
\usepackage{graphicx}
\usepackage{epsfig}
\usepackage{amsmath,amssymb}
\usepackage{lscape}
\usepackage{verbatim}
\usepackage{rotating}
\usepackage{epstopdf}
\usepackage{rotating}
\usepackage{pdflscape}
\usepackage{subfigure}
\usepackage{chngcntr}
\usepackage{float}
\usepackage{placeins}
\usepackage{color}

\def\lapprox{\lower.4ex\hbox{$\;\buildrel <\over{\scriptstyle\sim}\;$}}
\def\gapprox{\lower.4ex\hbox{$\;\buildrel >\over{\scriptstyle\sim}\;$}}

\DeclareRobustCommand{\ion}[2]{%
\relax\ifmmode
\ifx\testbx\f@series
{\mathbf{#1\,\mathsc{#2}}}\else
{\mathrm{#1\,\mathsc{#2}}}\fi
\else\textup{#1\,{\mdseries\textsc{#2}}}%
\fi}

\title[The Hot Halo Around NGC1961]{A Deep XMM-Newton Study of the Hot Gaseous Halo Around NGC 1961}
\author[Anderson, Churazov, and Bregman]{Michael E. Anderson$^{1}$\thanks{email: michevan@mpa-garching.mpg.de}, Eugene Churazov$^{1,2}$, Joel N. Bregman$^{3}$ \\
$^{1}$Max-Planck Institute for Astrophysics, Garching bei Muenchen, Germany\\
$^{2}$Space Research Institute (IKI), Profsoyuznaya 84/32, Moscow 117997, Russia\\
$^{3}$Department of Astronomy, University of Michigan, Ann Arbor, MI, USA}

\begin{document}

\maketitle

\begin{abstract}
We examine 11 XMM-Newton observations of the giant spiral galaxy NGC 1961, allowing us to study the hot gaseous halo of a spiral galaxy in unprecedented detail. We perform a spatial and a spectral analysis; with the former, the hot halo is detected to at least 80 kpc and with the latter its properties can be measured in detail up to 42 kpc. We find evidence for a negative gradient in the temperature profile as is common for elliptical galaxies. We measure a rough metallicity profile, which is consistent with being flat at $Z \sim 0.2 Z_{\odot}$. Converting to this metallicity, the deprojected density profile is consistent with previous parametric fits, with no evidence for a break within 42 kpc ($\sim$0.1R$_{\text{vir}}$). Extrapolating to the virial radius, we infer a hot halo mass comparable to the stellar mass of the galaxy, and a baryon fraction from the stars and hot gas of around 30\%. The cooling time of the hot gas is orders of magnitude longer than the dynamical time, making the hot halo stable against cooling instabilities, and we argue that an extended stream of neutral Hydrogen seen to the NW of this galaxy is instead likely due to accretion from the intergalactic medium. The low metallicity of the hot halo suggests it too was likely accreted. We compare the hot halo of NGC 1961 to hot  halos around isolated elliptical galaxies, and show that the total mass determines the hot halo properties better than the stellar mass.  
\end{abstract}
 

\section{Introduction}

X-ray emission from hot gas appears to be a generic feature of massive halos. Hot gaseous halos suffuse nearly every galaxy cluster and many galaxy groups (\citealt{Forman1982}, \citealt{Sarazin1986}, \citealt{Sun2009}, \citealt{Kravtsov2012}), and they are also very common (possibly ubiquitous) around massive elliptical galaxies (\citealt{Forman1985}, \citealt{Fabbiano1989}), including field ellipticals \citep{Anderson2015}. The hot halo is strongly affected by both feedback from the galaxy and by accretion from the intergalactic medium (e.g. \citealt{Cen2006}, \citealt{Roncarelli2012}). These processes, which are fundamental for understanding galaxy formation, can therefore be studied through X-ray observations of the hot halos.

In the aggregate, X-ray emission can be described by power-law relations as a function of stellar mass (e.g. \citealt{Helsdon2001}, \citealt{OSullivan2003}, \citealt{Mulchaey2010}, \citealt{Boroson2011}) and halo mass (\citealt{Kaiser1986}, \citealt{Reiprich2002}), although the scatter in these relations is considerable. Potentially active galactic nucleus (AGN) feedback (e.g. \citealt{Churazov2000}, \citealt{Churazov2001}) might also be important. For example, the slopes of these power-law relations differ from the self-similar prediction, which can be used as a constraint on the effect of AGN feedback (\citealt{Gaspari2014}, \citealt{Anderson2015}). For clusters and groups the luminosity and temperature of the gas are also related (e.g. \citealt{Mitchell1977}, \citealt{Mushotzky1978}, \citealt{David1993}, \citealt{Bryan1998}, \citealt{Pratt2009}), but this relation seems to break down at the scale of galactic halos (\citealt{Fukazawa2006}, \citealt{Humphrey2006}, \citealt{Diehl2008}). Some of the scatter in these relations may be correlated with the degree of rotational support in the central galaxy as well (\citealt{Sarzi2013}, \citealt{Kim2013}). 

Another important issue is the baryon fraction of massive halos. In galaxy clusters, the hot halos contain the overwhelming majority of the baryons associated with the cluster (\citealt{Ettori2003}, \citealt{Gonzalez2007}, \citealt{Dai2010}, \citealt{Lagana2013}), although the stellar phase begins to grow in importance as the halo mass decreases. An open question is the contribution of the gas phase in group and galaxy halos. If it is sufficiently massive, then the baryon fraction remains constant from galaxy clusters (which are nearly baryon-complete; \citealt{White1993}) down to isolated galaxies. Current censuses appear to show a declining baryon (stars + ISM + hot gas) fraction as the halo mass decreases (\citealt{McGaugh2010}, \citealt{Anderson2010}, \citealt{Papastergis2012}), suggesting a problem of ``missing baryons'' from galaxies, but significant questions remain. 

Studies by \citet{Planck2013} and \citet{Greco2014} find a self-similar relation as a function of halo mass for the hot gas pressure as probed by the Sunyaev-Zel'dovich (SZ) effect. This suggests that, at least within the Planck beam (typically several galactic virial radii), the Cosmic baryon fraction is approximately recovered in hot gas (although see \citealt{LeBrun2015}).  There are also claims of individual elliptical galaxies with extremely massive gaseous halos that bring the baryon fractions of the systems up to the cosmological value (\citealt{Humphrey2011}, \citealt{Humphrey2012}), although for at least one of these claims it has been shown that the spectral modeling employed for the analysis violates X-ray surface brightness constraints and therefore significantly overestimates the total hot gas mass \citep{Anderson2014}.  Many simulations predict that the density profile of the hot halo should systematically flatten as the halo mass decreases, due to the increasing importance of galactic feedback which inflates the entropy of the hot halo (\citealt{White1991}, \citealt{Benson2000}). A flatter density profile would have less pronounced X-ray emission, potentially reconciling the SZ results with the X-ray constraints.
 
Around spiral galaxies, the picture is a bit different. Starbursting galaxies produce X-ray emitting winds (\citealt{Strickland2000}, \citealt{Strickland2004}, \citealt{Tullmann2006}, \citealt{Li2013}), but extended gaseous halos have generally proven to be very difficult to detect around spiral galaxies (\citealt{Bregman1982}, \citealt{Benson2000}, \citealt{Rasmussen2009}, \citealt{Bogdan2015}). The exceptions are the most massive spiral galaxies. Extensive hot halos have been detected around the giant spirals NGC 1961, UGC 12591, NGC 6753, and 2MASX J23453268-0449256 (\citealt{Anderson2011}, \citealt{Dai2012}, \citealt{Bogdan2013}, \citealt{Walker2015}).  There is also a weak ROSAT detection of hot gas around NGC 266 \citep{Bogdan2013b}, and suggestion of extended hot gas around a stack of ROSAT images of nearby isolated spirals \citep{Anderson2013}. 

It is unclear what explains the difficulty of detecting hot gaseous halos around massive spiral galaxies. Weak lensing studies (e.g. \citealt{Sheldon2004}, \citealt{Hoekstra2005}, \citealt{Mandelbaum2006}, \citealt{Velander2014}) generally find that massive blue centrals and massive red centrals obey different  $M*-M_{\text{halo}}$ relations, so that at fixed stellar mass, massive blue galaxies lie in less massive halos than red galaxies. If the gaseous halo is responding primarily to the potential of the dark matter halo, this could plausibly explain the relative X-ray faintness of hot halos around spiral galaxies. Alternatively, if the hot halo is tightly coupled to the galaxy (through feedback, accretion, or some combination of the two), then the differences in feedback and accretion between spirals and ellipticals might explain the different hot halo properties. 

Around the massive spirals, the hot gas generally appears to be smoothly distributed with approximate azimuthal symmetry, and has been detected to radii of 60 kpc or so from the galaxy (although there is no indication that the halo truncates here). The temperatures of the hot halos are generally close to the expected virial temperatures of these systems, suggesting the hot gas is roughly in hydrostatic equilibrium with the halo potential. Comparisons with cosmological simulations have been able to reproduce the X-ray emission around spirals  (\citealt{Li2014}, \citealt{Bogdan2015}), although these comparisons have been unsuccessful in reproducing the X-ray properties of spirals and ellipticals simultaneously. A unified picture of the X-ray circumgalactic medium is still lacking. It is particularly important to push the X-ray observations of spiral galaxies to larger radii, to probe the regime where most of the mass in the halo is predicted to lie, and to be able to constrain the properties of the dark halo and the hot halo simultaneously. 

The real quantities of interest are the radial density and temperature profiles of the gas (or equivalently the entropy and pressure profiles), since these encode the feedback history of the galaxy and determine the behavior of the hot halo. The temperature profile can easily be measured from the X-ray spectrum, assuming enough photons are available to divide the field into several annuli. So far no observations have been deep enough to permit this measurement for the hot halo of a spiral galaxy.

The density profile can be inferred from the X-ray surface brightness profile, using an assumed gas temperature and metallicity profile and a few assumptions to deproject the observations. Typically spherical symmetry and a flat abundance profile are assumed, as well as either a flat temperature profile (leading to a ``beta profile'' for the density; \citealt{Cavaliere1978}) or a temperature profile appropriate for a cool-core galaxy cluster (leading to a modified ``beta profile'' for the density; \citealt{Vikhlinin2006}). The assumed metallicity has a significant effect on the final result, since the plausible range of gas metallicities spans more than an order of magnitude (roughly 1/10 Solar to slightly super-Solar). For metallicities in this range, the conversion from soft-band surface brightness to density depends approximately on the square root of the gas metallicity. In terms of the total gas mass, the extrapolation to large radii is even more significant since the hot halo is typically only detected out to about a tenth of the virial radius; most of the inferred gas mass lies at larger radii where we have few constraints on the density profile. The farther out the surface brightness profile can be measured, the more reliable the extrapolation becomes, leading to better estimates of the total hot halo mass. 

In order to improve the constraints on the density profile, and to make a first measurement of a temperature profile, we re-observed the giant spiral galaxy NGC 1961 with XMM-Newton for an additional 215 ks (adding to the 74 ks of observations of this galaxy already taken with XMM-Newton and presented in \citet{Bogdan2013}). In this paper we report our analysis of these observations of NGC 1961. In section 2 we discuss the data reduction. In Section 3 we present a spatial analysis of this galaxy, and in Section 4 we present a spectral analysis. In section 5 we combine these analyses and measure the pressure and entropy profiles of the hot halo. In section 6 we discuss our results in the context of the ``missing baryons'' problem and in comparison to isolated elliptical galaxies.

NGC 1961 is an extremely massive late-type spiral galaxy. Based on its recessional velocity of 3934 km/s listed in the NASA Extragalactic Database, and an assumed Planck cosmology \citep{Planck2015} with $H_0$ = 67.8 km/s/Mpc, we estimate the distance of NGC 1961 to be 58.0 Mpc, so that one arcminute corresponds to 16.6 kpc. The K-band luminosity of this galaxy is then $5.6\times10^{11} L_{\odot}$, corresponding to a stellar mass of $3\times10^{11} M_{\odot}$ (assuming a M/L ratio of 0.6, which is the rough expectation based on its K-band luminosity; \citealt{Bell2001}). For the B-V$=0.6$ color of this galaxy, a \citet{Chabrier2003} initial mass function (IMF) gives the same M/L ratio, but a \citet{Salpeter1955} mass function gives a M/L ratio of 1.2, yielding an even larger stellar mass for this galaxy. However, this latter M/L ratio is disfavored by \citet{McGaugh2015}, who use both population synthesis and IMF-independent constraints to determine a nearly universal M/L ratio of 0.57 in the K-band, close to our value of 0.6. The inclination-corrected HI rotation velocity is 340 km/s (Haan et al. 2008) at a projected distance of 43 kpc (although it reaches higher values -- up to about 450 km/s -- at smaller radii). The virial radius of this galaxy is approximately 490 kpc, as we discuss in section 5.1. We use the coordinates from the NASA Extragalactic Database for the center of the galaxy.

\section{Observations and Data Reduction}

We obtained nine observations of NGC 1961 with XMM-Newton, ranging from 22-27 ks in length each. The aimpoints of each observation were varied such that they formed a $3\times3$ grid, with a typical separation between aimpoints of about an arcminute. This layout makes the data reduction and spectral fitting more cumbersome, but it offers the advantages of flattening out the exposure map around the galaxy and reducing the effects of vignetting. It also allows us to identify and separate various non-astrophysical background components (soft protons, instrumental background, solar wind charge exchange) from the astrophysical backgrounds like the Galactic halo and the Cosmic X-ray background. We also re-analyzed the two previous XMM-Newton observations of this galaxy which were originally discussed in \citet{Bogdan2013} (hereafter B13). The list of all 11 observations is presented in Table 1. 

\begin{table}
\begin{centering}
\caption{XMM-Newton Observations of NGC 1961}
\begin{tabular}{ccccc}
\hline
obsid  &  $T_{\text{exp}}$&   $T_{\text{MOS1}}$ & $T_{\text{MOS2}}$ & $T_{\text{PN}}$\\
 & (ks) & (ks) & (ks) & (ks) \\
\hline
0673170101 & 37.9 & 20.4 & 20.5 & 15.2 \\
0673170301 & 35.9 & 13.1 & 16.8 & 7.0 \\
0723180101 & 23.7 & 18.8 & 19.2 & 14.0 \\
0723180201 & 22.9 & 5.4 & 5.9 & 2.4 \\
0723180301 & 26.5 & 9.5 & 10.5 & 2.4 \\
0723180401 & 23.7 & 3.2 & 6.8 & 1.6 \\
0723180501 & 22.9 & 2.7* & 4.0* & 1.6* \\
0723180601 & 26.9 & 6.6 & 8.1 & 4.8 \\
0723180701 & 22.0 & 7.4 & 7.1 & 5.3 \\
0723180801 & 22.0 & 11.9 & 12.4 & 8.3 \\
0723180901 & 24.5 & 10.0 & 9.5 & 5.6 \\
\hline
total & 289.0  & 106.5 & 116.8 & 65.6\\
\hline
\end{tabular}
\\
\end{centering}
\small{List of XMM-Newton observations of NGC 1961. The first column is the obsid and the second column is the total duration of the observation as listed in the XMM-Newton Science Archive. The next three columns show the length of the good time intervals (GTIs) for each instrument after using the XMM-ESAS software to filter each observation, as described in the text. These observations were heavily contaminated by flaring; much less than half of the total exposure time was useable for analysis. Observation 0723180501 was so heavily contaminated that it was still not useable even after GTI filtering, so we discard this observation from subsequent analysis (and do not include it in the GTI filtered total exposure time at the bottom of this Table). }
\end{table}

We reduced the data following the procedure outlined in the XMM-Extended Source Analysis Software (ESAS) Cookbook \citep{Snowden2014}. We used HEASOFT v. 6.16 and XMM-SAS v. 13.5.0 for the data reduction and analysis. We first ran the initial processing commands (\verb"epchain", \verb"epchain withoutoftime=true", \verb"pn-filter", \verb"emchain", \verb"mos-filter") to produce filtered events files for each observation. The data were heavily affected by flaring, so this processing was especially important. We examined each lightcurve manually to verify that the pipeline processing was working correctly. We also examined the filtered events files for CCDs in anomalous states and excluded them. During this process we found that obsid 0723180501 was so heavily affected by flaring that it was effectively unusable; we therefore discarded this observation for the subsequent analysis. Table 1 shows the effects of the initial processing on each event file. 

Next, we identified point sources and produced point source masks for each observation. Since the fields of view overlap in each observation, we produced broad-band images and exposure maps from each events file and merged them into a single broad-band image. We then ran the Chandra \verb"wavdetect" algorithm on the merged image with default parameters in order to identify point sources; at a projected radius of 3' (50 kpc) from the center of the galaxy we estimate our limiting point source flux to be around $7\times10^{-17}$ erg s$^{-1}$ cm$^{-2}$ in the soft (0.4-1.25 keV) band, corresponding to a luminosity of $3\times10^{37}$ erg s$^{-1}$. This is sufficient to resolve the brightest X-ray binaries, but there will be an important unresolved component remaining. We verified the point source list with manual inspection, then passed the list of sources to the \verb"region" and \verb"make-mask" routines in order to construct masks with the appropriate radii for each observation (setting the radius to encircle 85\% of the expected energy for each point source).

Finally, we used the \verb"mos-spectra", \verb"mos_back", \verb"pn-spectra", and \verb"pn_back" routines to generate broad-band spectra and to generate images in the 0.4-1.25 keV (``soft'') and 2.5-7.0 keV (``hard'') bands. These routines also generate background files (spectra and images) with the estimated particle background in each observation. We bin these spectra such that there are at least 20 counts in each energy bin.

To fit the remaining backgrounds, we extract a spectrum from each observation after masking out point sources and masking out a circle centered on the galaxy with radius 8' in order to exclude any possible emission from the galaxy or its hot halo. We bin each of these background spectra as above, and perform a joint analysis of all 30 spectra using XSPEC v. 12.8.2 \citep{Arnaud1996} and the associated Python wrapper PyXSPEC. We analyze the MOS spectra in the 0.3-11.0 keV energy band and the PN spectra in the 0.4-6.5 keV energy band (at higher energies the PN spectra are affected by a number of instrumental lines). We use the  abundance tables of \citet{Anders1989}.

The spectral model is complex and is based on the proposed model outlined in the ESAS cookbook. We use two Solar abundance APEC models \citep{Smith2001} with temperatures of 0.1 and 0.25 keV to model the Local Bubble and the Galactic halo respectively. We model the Cosmic X-ray background (CXB) with an absorbed $\Gamma = 1.44$ powerlaw, and we set the absorption to the Galactic value at the location of NGC 1961 ($8.2\times10^{20}$ cm$^{-2}$, \citealt{Dickey1990}, \citealt{Kalberla2005}). These components each have free normalization, but we tie the normalizations in each observation together (with appropriate correction factors for the differences in the area of each region, and with a free parameter which is allowed to range from 0.9-1.1 for each observation to allow for uncertainties in the calibration, as suggested by \citet{Snowden2014}). We also include a number of components in our model which account for instrumental or time-variable backgrounds, and are therefore allowed to very between observations. First is a zero-width Al K$\alpha$ instrumental line at 1.49 keV with free normalization for each detector and each observation, and a zero-width Si K$\alpha$ instrumental line at 1.75 keV with free normalization for the MOS detectors in each observation. Second is a power-law for the soft proton background, with free normalization and slope for each detector and each observation, although the slope is constrained to have an index between 0.1 and 1.4. This component uses a diagonal response matrix provided with the XMM-SAS software package. The ESAS cookbook suggests that one may assume the soft proton background in each observation has the same slope for both MOS detectors, but given the potential importance of the soft proton background due to the significant flaring in our observations, we choose to relax this assumption and fit the proton background for each instrument separately. Finally, we also include six zero-width line features to account for the Solar wind charge exchange (SWCX) background. These lines have energies fixed at 0.46, 0.57, 0.65, 0.81, 0.92, and 1.35 keV corresponding to \ion{C}{vi}, \ion{O}{vii}, \ion{O}{viii}, \ion{O}{viii}, \ion{Ne}{ix}, and \ion{Mg}{xi} transitions respectively. The SWCX background is allowed to vary between observations but its normalization is fixed across all three instruments during each observation. 

In order to reduce the dependence of the results on initial guesses, we used the \verb"steppar" command for each of the SWCX lines to explore different values for the normalization and improve the fits. The final fit had a reduced $\chi^2$ of $1.0134$ for 8615 degrees of freedom, which is an acceptable fit. We use the normalizations of the CXB, Local Bubble, Galactic halo, and SWCX lines for modeling the background in the subsequent analysis, as well as the slope and normalization of the soft proton component.

\section{Spectral Analysis}

For the first time, we have enough X-ray photons to perform spectral fitting in multiple annuli around an isolated spiral galaxy.  We define nine concentric regions, whose layout is illustrated in Figure 4, along with an optical image of the galaxy. The first region is a circle of radius one arcminute centered at the nucleus of NGC 1961. The other eight regions are annuli of width one arcminute, centered at projected radii of 1'.0, 1'.5, 2'.0, 2'.5, 3'.0, 3'.5, 4'.0, and 4'.5 from the nucleus of the galaxy. Each region therefore overlaps with the adjacent regions. We chose this layout to improve the spatial resolution of our spectral analysis, and to test the robustness of our spectral modeling (i.e. overlapping regions should give consistent spectral results). However, the overlap means that adjoining spectra are not independent of one another, which we show graphically in subsequent plots through the X error bars spanning the radial range covered by each spectrum. We refer to these these regions as R0 through R8, with the number increasing with projected radius. R0 covers the majority of the galaxy and its bulge, R1 is dominated by the disk, and R2 by the outer disk. The other six regions capture the hot halo.

\begin{figure*}
\begin{center}
\subfigure[Regions R0, R2, R4, R6, and R8]{\includegraphics[width=8cm]{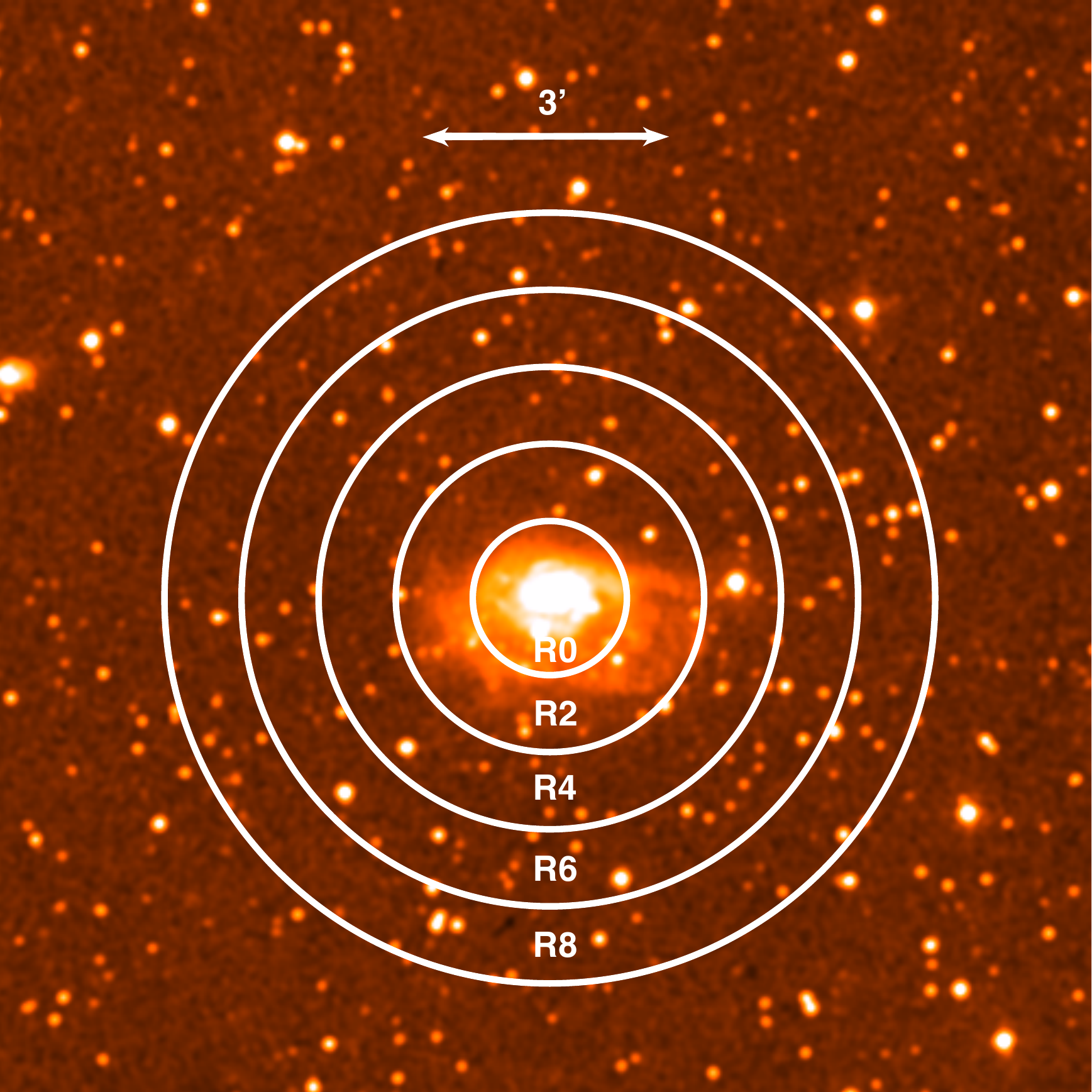}}
\subfigure[Regions R1, R3, R5, and R7]{\includegraphics[width=8cm]{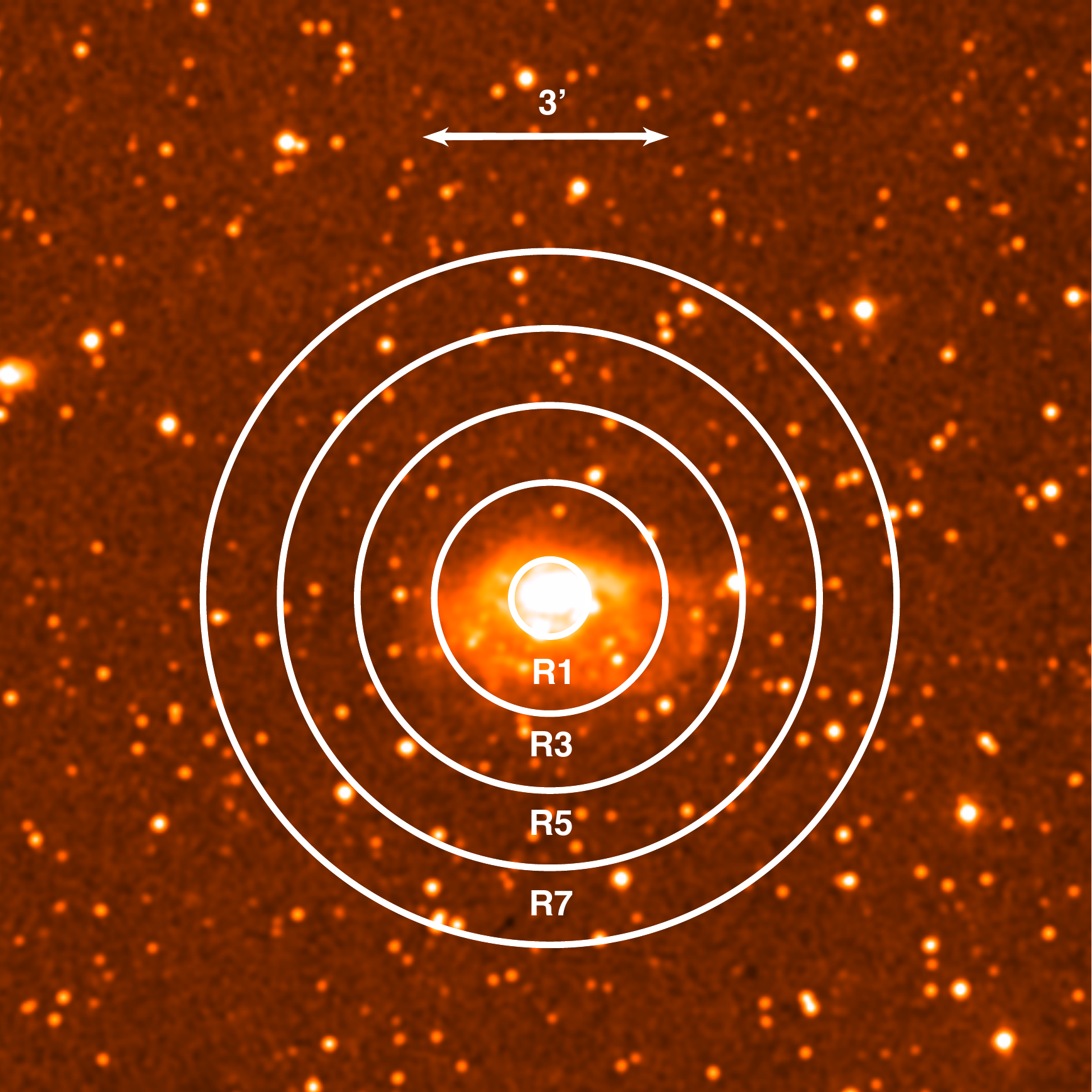}}
\end{center}
\caption{An optical DSS image of NGC 1961 (the same image is shown in both panels), along with our nine primary spectral extraction regions. Each region has a width of 1' (16.6 kpc). Note that regions R1, R3, R5, and R7 (shown in right) overlap with regions R0, R2, R4, R6, and R8 (left). The first region (R0) covers the majority of the galaxy and its bulge, while the outer disk dominates region R2. Regions R3-R8 capture the emission from the hot halo. }
\end{figure*}

We extract the spectra from each of these nine regions in each of our 30 observations. For each region, we perform a joint fit to these 30 spectra using the model described in section 2, with an additional set of components for the source emission. We use a redshifted photoabsorbed APEC + powerlaw to describe the source emission. We fix the redshift to the value for NGC 1961 ($z = 0.013$). We try to fix as many free parameters as possible before performing the spectral fitting. We use the fits to the spectrum in the background region to get many of these constraints. The normalizations of the SWCX lines, the CXB, the Local Bubble component, and the Galactic halo are all fixed to the values obtained from the fit to the background region (scaled appropriately for each observation to account for the differences in angular area). In our fiducial model, we also freeze the X-ray binary (XRB) component, the soft proton component, and the metallicity of the APEC component. We now discuss each of these choices in detail.

The X-ray binary component is modeled as a powerlaw with redshifted photoabsorption from NGC 1961 as well as Galactic absorption. In our fiducial model we freeze the slope of this powerlaw at a reasonable value of $\Gamma=1.56$ \citep{Irwin2003}. For this model we also set constraints on the normalization of this powerlaw, by examining the surface brightness profile of the hard-band image (see next section). In the hard-band image, the hot halo of NGC 1961 is expected to contribute negligible emission, so we can attribute all the observed emission to XRBs. We measure the background-subtracted hard-band flux in each region and use this flux (and an assumed $\Gamma=1.56$ powerlaw spectral shape) to estimate the expected soft-band flux. This conversion is slightly uncertain, primarily due to uncertain background subtraction in the hard-band image and due to unknown absorption from NGC 1961 (since we do not constrain the absorption from NGC 1961 until performing the spectral fitting), so we allow the normalization of the XRB component to vary slightly around the expected value in our fiducial model. 

In order to test that our modeling of the XRBs does not introduce any systematic effect, we also explore an alternate set of spectral models where we relax the above constraints. We let the normalization float as a free parameter. In region R0, we also let the slope vary (between $\Gamma=1.2$ and $\Gamma=2.0$), and we tend to find that a harder slope than $\Gamma=1.56$ is preferred, probably due to a low-luminosity AGN. In regions R1 and R2, we let the slope vary between $\Gamma=1.5$ and $\Gamma=2.0$ to allow for the presence of high-mass X-ray binaries (which are typically softer than low-mass X-ray binaries). In the other regions, where very little XRB emission is expected, we fix $\Gamma=1.56$; in these regions the XRB component is typically unimportant, and the data are insufficient to constrain the slope.

The soft proton component is also modeled with a powerlaw, though this powerlaw is not folded through the standard instrumental response. We freeze the slope of this powerlaw to the slope measured in the background region (as \citealt{Snowden2014} note, it is generally reasonable to assume the spectral shape of the proton background is spatially constant across a given observation). In our fiducial model we also freeze the normalization of the soft proton component, using the ESAS \verb"proton_scale" routine in order to convert the normalization from the measured value in the background region into the appropriate value for each region for each observation. Again, in order to ensure that our constraints on this component do not systematically affect our conclusions, we also explore an alternate set of spectral models where we relax the above constraints. In this case, we relax them by allowing the normalization of the soft proton background to vary by up to 100\% in each observation, while keeping the slope fixed at the fiducial value. 

Finally, the metallicity is a key parameter of interest, but this is a very difficult quantity to measure observationally. We discuss this parameter in more detail below (in section 3.1), but in brief the difficulties arise due to a degeneracy between the metallicity and the normalization. In our fiducial model, we assume an intermediate value for the metallicity ($Z = 0.5 Z_{\odot}$), which reduces the magnitude of the possible error on the normalization (and the inferred density of the hot gas). To first order, our results can be scaled for a different assumed value for the metallicity by multiplying the density by the inverse square root of the fractional change in metallicity. However, again, we want to check that our assumption for the metallicity does not introduce a systematic error on the other components, so we explore an alternate set of spectral models with the abundance as a free parameter.

We therefore have three sets of constraints we can toggle for our spectral modeling, yielding eight total spectral models. The fiducial model uses all three of these constraints, and the other seven models relax one, two, or three of these constraints. The dispersion between the results obtained from fitting with each different model gives us a way to estimate the systematic uncertainty in our results due to the use of simple spectral models for fitting to a complex astrophysical system. 

Finally, after finding the best-fit parameters for each model in XSPEC, we use the \verb"chain" command to perform a Markov Chain Monte Carlo (MCMC) search of the parameter space in order to account for degeneracies between parameters and to make sure we properly sample the multidimensional space. XSPEC has two implementations of MCMC algorithms -- Goodman-Weare and Metropolis-Hastings -- and we explored both but due to the hard limits on the scaling factors for each instrument (allowed to range from 0.9-1.1) we achieved faster convergence and a higher acceptance fraction using the Metropolis-Hastings algorithm\footnote{During our analysis, XSPEC was updated to introduce changes to the operation of the MCMC chain command. The MCMC results in this work have been obtained using the newest version of XSPEC (v. 12.8.2q).}. We initialized the chain using a diagonal covariance matrix with covariances taken from the XSPEC fits.  We burn $10^4$ elements before running the chain for $10^5$ iterations, and apply a simple ``simulated annealing'' prescription for reducing the Metropolis-Hastings temperature of the fit as it progresses (we use an initial temperature of 5, and every 5000 steps we reduce the temperature by a factor of 0.9). This samples the parameter space more fully before the chain begins to converge. We take the median value of each parameter (marginalized over the others) as the best-fit value and we also record the central 90\% confidence interval of each parameter in order to quantify the uncertainties. Some of these values are listed for the fiducial model in Table 2 below, for the regions where the emission measure of the hot gas component is at least 3$\sigma$ above zero (note that the uncertainties listed in Table 2 bound the 90\% central confidence interval, and are therefore larger than the 1$\sigma$ uncertainties). 

\begin{table*}
\begin{centering}
\begin{minipage}{90mm}
\caption{Spectral Fitting Results}
\begin{tabular}{clllll}
\hline
region  & $kT$  & $\int n_en_HdV$ & Area & $\chi^2$ $/$ d.o.f. \\
 & (keV) &  ($10^{62}$ cm$^{-3}$) & (arcmin$^2$) &    \\
\hline
R0 & $0.73_{-0.02}^{+0.02}$ & $26.3^{+1.1}_{-0.9}$& 2.8& 1026.6/587\\ 
R1 & $0.63^{+0.04}_{-0.03}$ & $16.2^{+1.1}_{-0.8}$ & 5.4& 646.7/471\\
R2 & $0.63^{+0.08}_{-0.08}$ & $10.1^{+1.4}_{-0.9}$ & 7.7& 380.7/311\\
R3 & $0.41^{+0.12}_{-0.07}$ & $8.0^{+2.1}_{-1.9}$ & 9.2& 418.8/368\\
R4 & $0.39^{+0.25}_{-0.08}$ & $4.7^{+2.0}_{-1.7}$ & 10.1& 362.9/373\\
R012 & $0.73^{+0.02}_{-0.02}$&$34.4^{+1.9}_{-1.2}$ & 10.5& 1122.8/819\\
R123 & $0.61^{+0.03}_{-0.03}$&$21.7^{+1.4}_{-1.1}$ & 14.6 & 1055.8/879\\
R234 & $0.50^{+0.12}_{-0.10}$&$14.5^{+3.8}_{-2.1}$&17.8 & 956.6/877\\
R345 & $0.37^{+0.12}_{-0.05}$&$10.7^{+2.2}_{-3.1}$ &21.7 & 1012.8/955\\
R456 & $0.36^{+0.28}_{-0.07}$&$5.9^{+2.4}_{-2.9}$ &26.3 & 1138.8/1097\\
\hline
\end{tabular}
\\
\small{Results for the hot gas component of the emission from NGC 1961, based on spectral fitting using our fiducial model. This model has the metallicity frozen at $Z = 0.5 Z_{\odot}$; for other choices of metallicity, the normalization should be scaled by a factor of approximately $\sqrt{0.5/Z}$. The values and quoted errors are based on our MCMC chains. The best-fit values are the medians of the chain and the uncertainties bound the 90\% central confidence region around the median. The area is determined from the \verb"BACKSCALE" parameter in the spectrum, and shows the angular area of the fiducial spectrum for each region; this can be used to convert the emission measure into an average electron density, as we do below. In each of the listed regions, the hot gas component of the spectrum is significant at more than 3$\sigma$. }
\end{minipage}
\end{centering}
\end{table*}

We also examine five larger annuli which are constructed by combining the annuli shown in Figure 1. These annuli are 2' in width, instead of the 1' annuli shown in Figure 1. These larger annuli have more photons and therefore yield better constraints, especially on the metallicity of the hot gas, but we find no systematic differences based on the different annuli sizes. 

In Figure 2 we present the resulting temperature profile, and in Figure 3 we present the metallicity profile. The black points (with 90\% confidence regions) are the results for the fiducial model, and we plot the median results for each of the other seven models as well. The radii outside of which the normalization of the hot gas component falls below 3$\sigma$ are indicated as hatched shaded regions; in these outer regions the spectra fitting is not reliable. This occurs at $r \gapprox$ 42 kpc for the 1' annuli. For the 2' annuli the hot halo is detected at $>$3$\sigma$ in every annulus.

\begin{figure*}
\begin{center}
\subfigure[1' annuli]{\includegraphics[width=8cm]{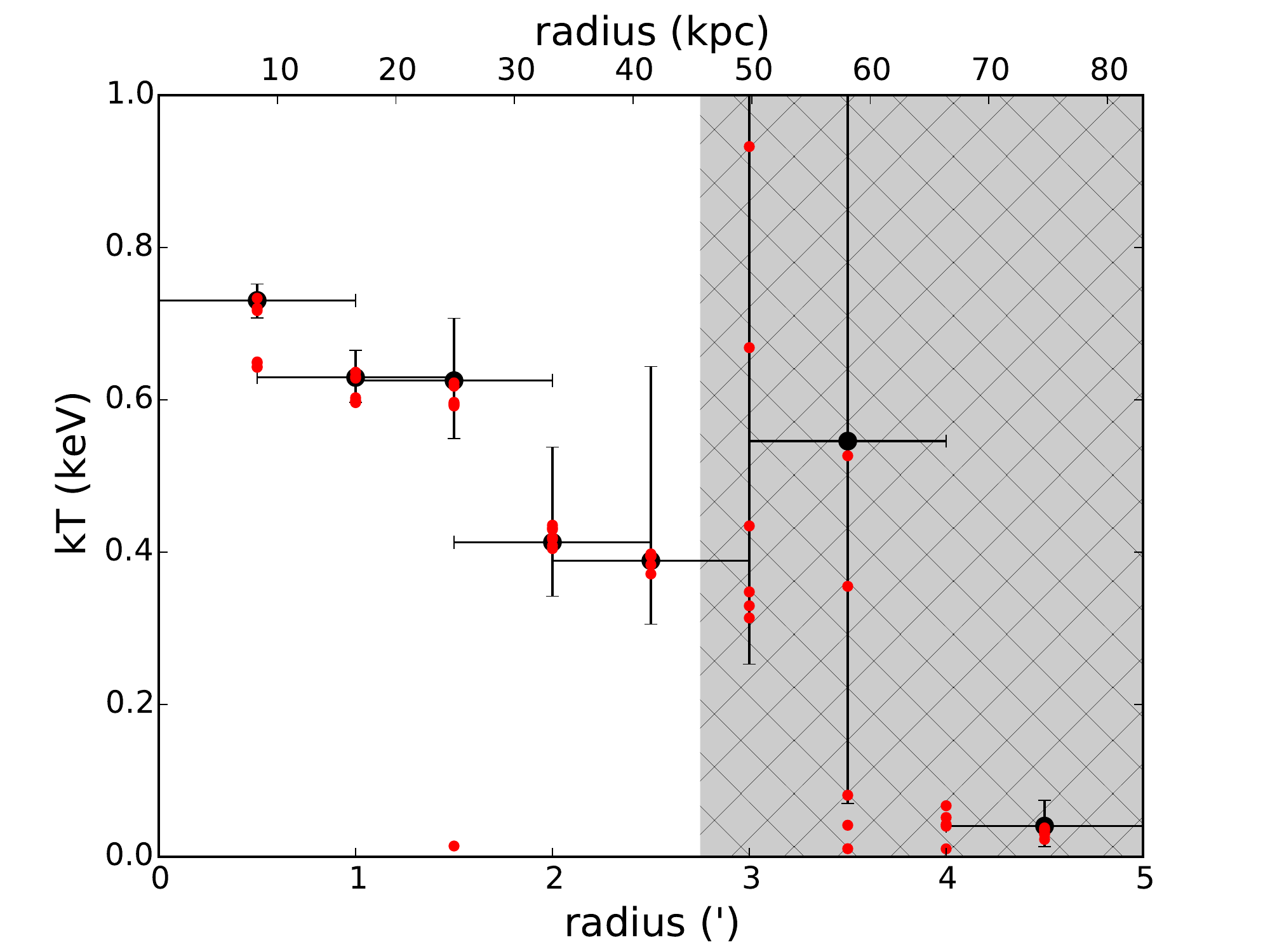}}
\subfigure[2' annuli]{\includegraphics[width=8cm]{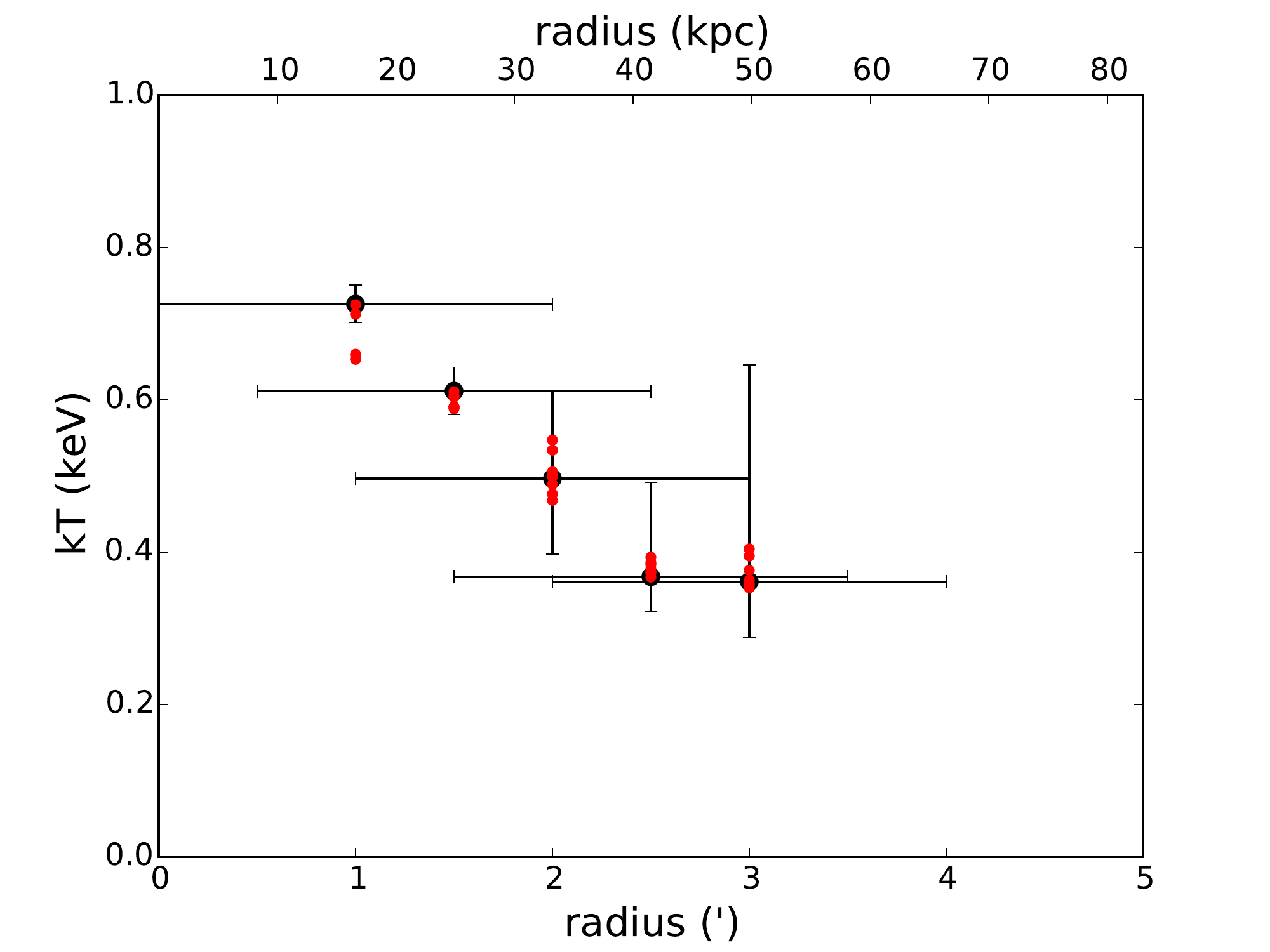}}
\end{center}
\caption{Temperature profile of the hot halo of NGC 1961, as measured using 1' annuli (left) and 2' annuli (right). Temperatures in each region are measured using eight different spectral models, as explained in Section 3. The best-fit values for the fiducial model, along with 90\% confidence intervals, are shown in black. The best-fit values for the other seven models are shown in red and give a sense of the systematic uncertainty stemming from the choice of models. Confidence intervals for the red points are similar in size to the confidence intervals for the fiducial model, but are not shown for clarity. For points in the shaded hatched region ($r \gapprox 42$ kpc), the hot gas component in the model has less than $3\sigma$ significance; results in this region are not reliable. The temperature appears to decline slowly with radius. }
\end{figure*}

\begin{figure*}
\begin{center}
\subfigure[1' annuli]{\includegraphics[width=8cm]{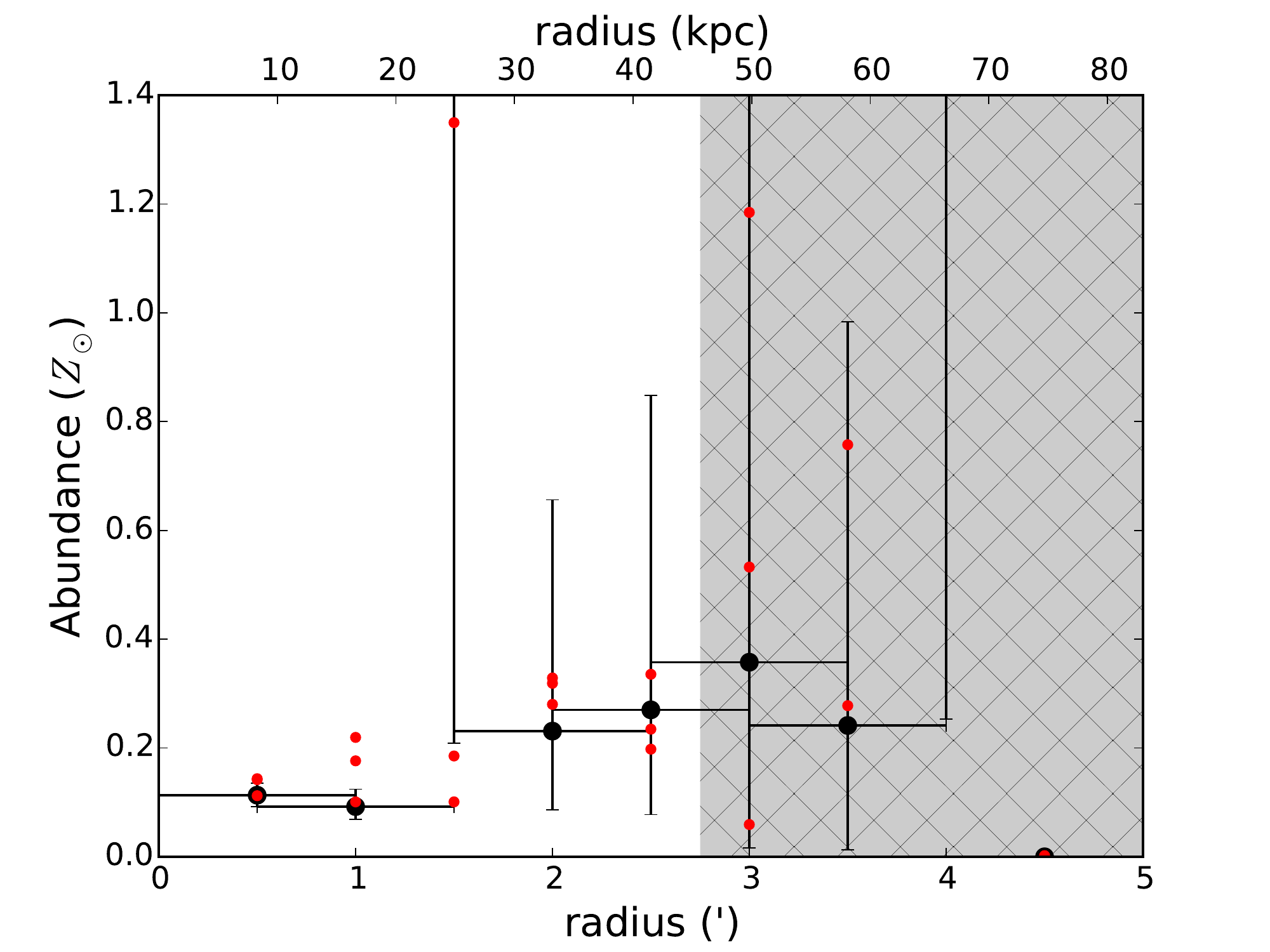}}
\subfigure[2' annuli]{\includegraphics[width=8cm]{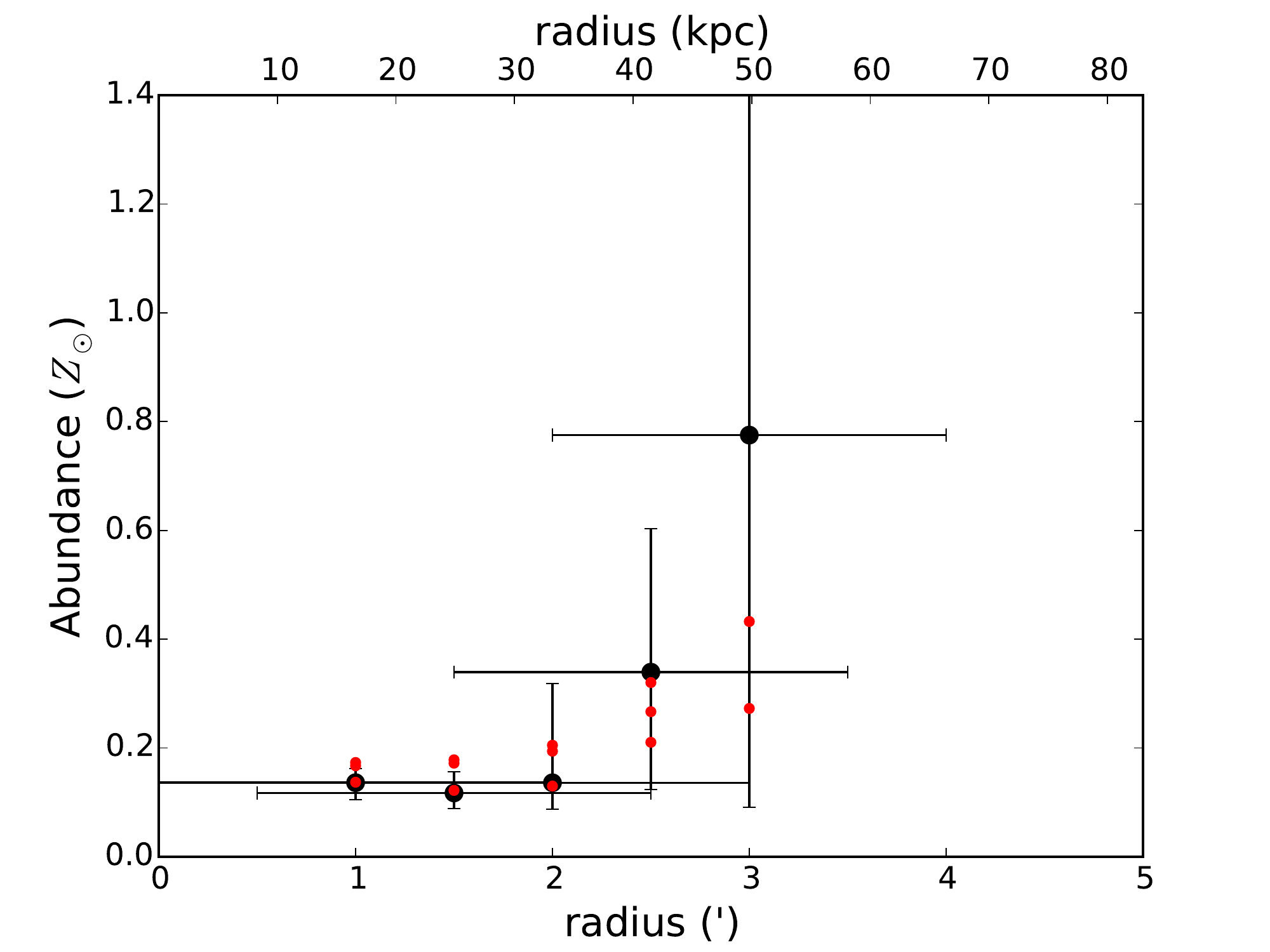}}
\end{center}
\caption{Metallicity profile of the hot halo of NGC 1961, as measured using 1' annuli (left) and 2' annuli (right). Temperatures in each region are measured using eight different spectral models, as explained in Section 3. The best-fit values for the fiducial model, along with 90\% confidence intervals, are shown in black. The best-fit values for the other seven models are shown in red and give a sense of the systematic uncertainty stemming from the choice of models. Confidence intervals for the red points are similar in size to the confidence intervals for the fiducial model, but are not shown for clarity. For points in the shaded hatched region ($r \gapprox 42$ kpc), the hot gas component in the model has less than $3\sigma$ significance; results in this region are not reliable. The metallicity is also essentially unconstrained in the 1' R2 annulus and in the outermost 2' annulus. Overall the profile is difficult to measure precisely, but it is consistent with being flat at a value around 0.2$Z_{\odot}$.}
\end{figure*}

The temperature in Figure 2 appears to decline slowly with radius. We can quantify the significance of this decline using the $\chi^2$ statistic, comparing a constant model to a linear model (so the two models differ by one degree of freedom). We fit to the spectral results for regions R0, R2, and R4, which are non-overlapping. We also include a systematic error on each temperature, in quadrature, which is derived from the standard deviation of the eight best-fit values for each region from the eight different models we consider. Even with the systematic error folded in, the linear model is highly favored, with a $\Delta \chi^2$ of 6.5 as compared to the flat model, giving 2.5$\sigma$ evidence for a negative temperature gradient. As an alternative metric, since the error bars are asymmetric and we have the full probability distributions from MCMC fitting, we also compute the Akaike information criterion for flat and linear models. This gives a similar result, favoring the linear (negative) profile at 97.7\% confidence (approximately 2.3$\sigma$). On the other hand, for the metallicity, both methods show that a linear model is not at all favored over a flat model at the present level of statistical accuracy, and we conclude the metallically profile is consistent with being flat.

\subsection{On the low value of the metallicity}

The metallicity profile in Figure 3 shows strong statistical evidence for sub-Solar metallicity throughout the hot gaseous halo.  While the hot gas abundance is observed to be sub-Solar in some X-ray faint elliptical galaxies (e.g. \citealt{Su2013}), metallicities as low as ours are still unusual, especially since a star-forming galaxy like NGC 1961 can be expected to have a higher supernova rate than a comparable elliptical. Such a low metallicity might suggest an external source (i.e. intergalactic medium) for the majority of the hot gas instead of an internal source. We think this result therefore warrants a bit more discussion. 

At temperatures around 0.6 keV, the metallicity is inferred from the ratio of the Fe L complex at around 0.7-0.9 keV to the pseudocontinuum at around 0.4-0.55 keV. This procedure breaks down when the metallicity becomes high enough that line emission begins to dominate over the continuum, so we performed simulations with XSPEC in order to verify that this is not a concern for these sub-Solar metallicities, and to verify that the uncertainties returned by our MCMC modeling are of the expected order for the numbers of photons in our spectra (approximately $10^4$ for the 1' regions and twice as high for the 2' regions). 

The reader can get a sense of the statistics in our spectra from Figure 4. For this figure, we have added the 20 MOS spectra for each region to generate composite spectra. {\bf We emphasize that we do not fit models to these composite spectra}; we fit to the individual spectra and propagate the backgrounds and angular areas separately for each spectrum. We show examples of the major model components in Figure 4 to illustrate their shape. The composite spectra show that the continuua are generally fairly well determined. The spectra behave roughly as expected as well, with the Fe L complex becoming increasingly weak as we move outwards in radius. Note also that we have not added the PN spectra to these composites; the PN spectra have roughly the same number of photons as the MOS spectra, improving our statistics by another factor of two.

\begin{figure*}
\begin{center}
\subfigure[Regions R0-R8]{\includegraphics[width=8cm]{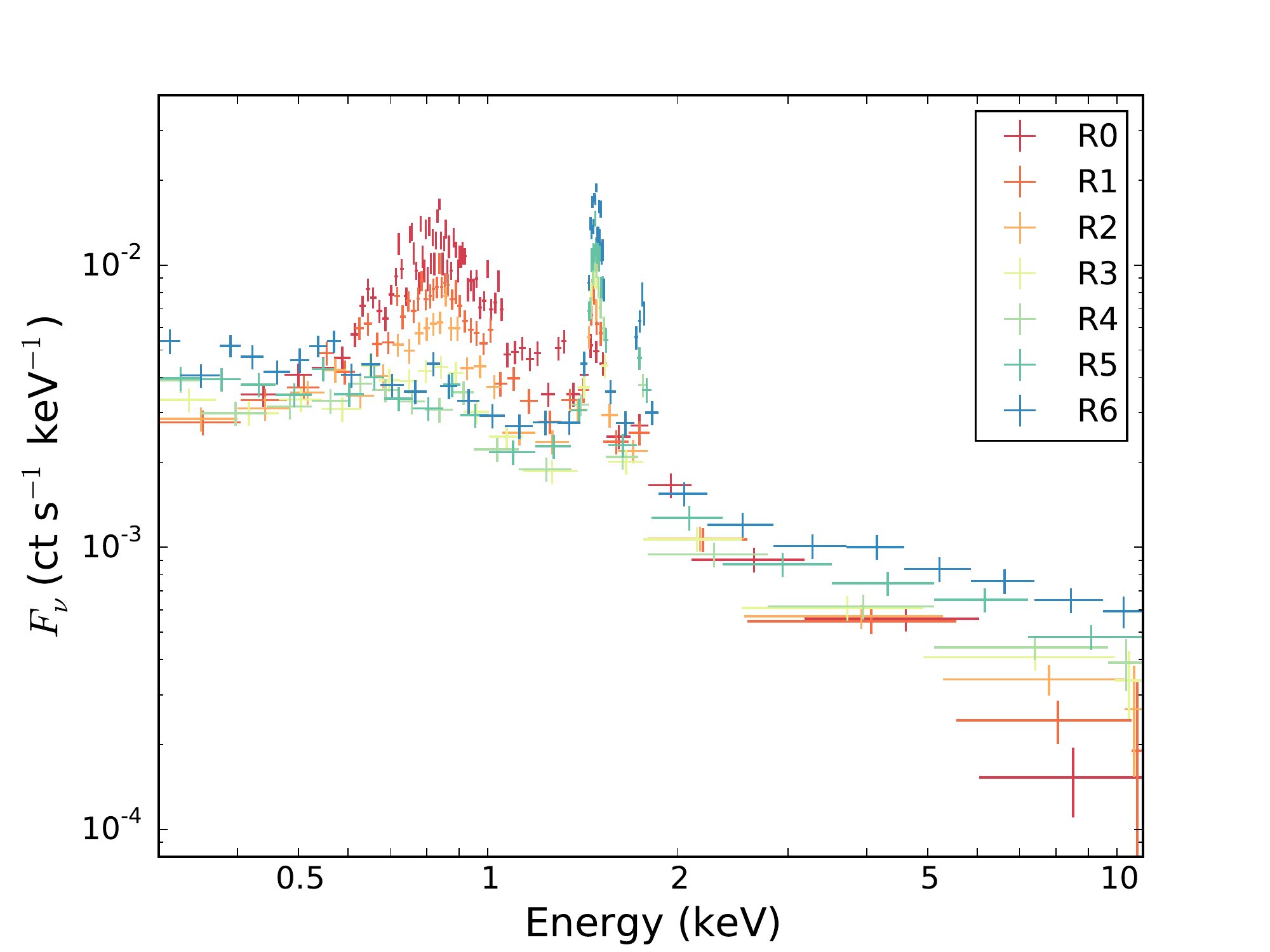}}
\subfigure[Region R2]{\includegraphics[width=8cm]{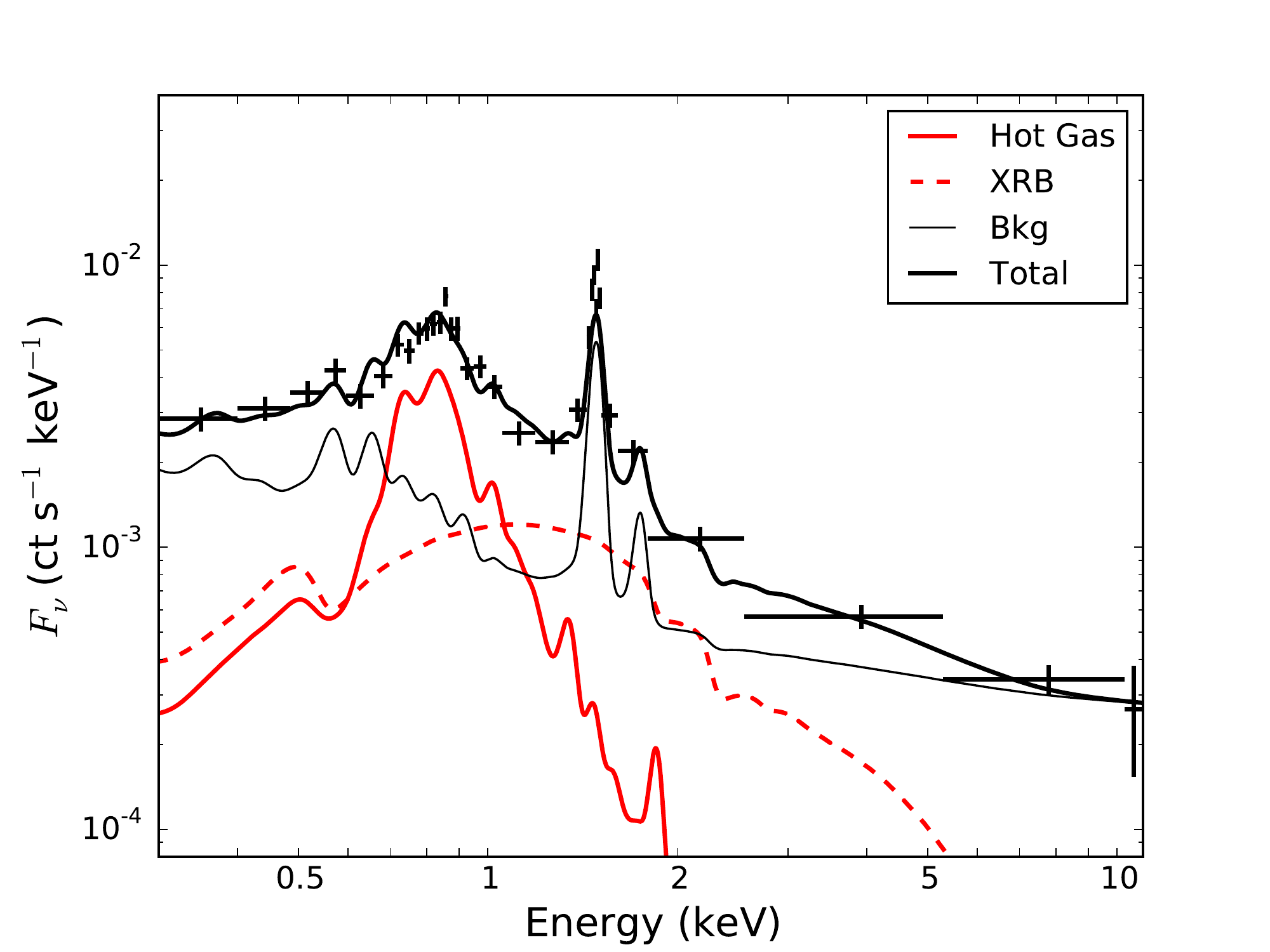}}
\end{center}
\caption{Illustrations of stacked MOS spectra from various regions. The left plot shows stacked MOS spectra from each region,  rebinned so that each point has a S/N of at least 10. Note that the hot gas component of the spectrum (roughly between 0.6 and 1 keV) becomes progressively less important at larger radii. In the right plot, we show the R2 spectrum as well as examples of representative spectral models for hot gas (red solid line), X-ray binaries (red dashed line), sky+instrumental background (thin black line), and the sum of these three components (thick black line). We stress that these stacked spectra and the models are for illustration purposes only. The actual spectral fitting is performed using simultaneous fitting to each of the 30 individual spectra for each region.}
\end{figure*}

There are potential systematic errors, however. One possibility is the tendency of single-temperature fits to multi-temperature plasmas to yield anomalously low metallicities \citep{Buote2000}, which we show in section 3.2 does not significantly affect our result. Another issue is the determination of the continuum, and our model contains many components so it is important to check whether any of them may influence our measurement of the continuum. As we will show in the spatial analysis, the hot gas is the dominant component of the soft-band flux within the inner two or three arcminutes, after which the sky background and the QPB become the largest and second-largest components, respectively. The QPB is mostly expressed through the bright instrumental lines, none of which lie near the 0.4-0.6 keV continuum, so we do not expect it to affect the inferred metallicity. The sky background is modeled with a combination of a power-law (for the CXB) and two APEC models (for the Galactic halo and the Local Bubble), each of which we fix based on the measurement in the background region, which begins 8' from the center of the galaxy. A fluctuation in these components on angular scales of several arcminutes would therefore lead to an incorrect assessment of the continuum in our source spectra. We can estimate the expected magnitude of the CXB fluctuations using the results of Kolodzig et al. (in prep) for the XBootes field. In the 0.5-2.0 keV band on 4' scales they find a typical size of CXB fluctuations of $5\times10^{-5}$ ct$^2$ s$^{-2}$ deg$^{-2}$ (Chandra ACIS-I counts). This corresponds to roughly 100 soft-band counts (XMM-EPIC MOS + PN instruments) in a 16 square-arcminute region observed for 100 ks, which is less than a percent of the counts in our spectra. It is possible that fluctuations between 0.4 and 0.5 keV are more significant, however, since they include more Galactic halo and Local Bubble emission, and the spatial variation of these components is not known. It is therefore not possible to firmly rule out the possible effect of background fluctuations from the CXB or Galactic backgrounds, although it would require significant fluctuations on these scales in order to bias our metallicity measurement. 

The SWCX and soft proton backgrounds also contribute to the 0.4-0.6 keV band. In the background region the SWCX component is about an order of magnitude below the CXB, however, and the SWCX background is not seen to express significant spatial variations within an XMM-Newton field of view \citep{Snowden2014}, so we do not think this component is likely to affect the metallicity measurement. The soft proton background is also similarly low in the background region, and if we were incorrectly measuring it in such a way as to affect the inference of the metallicity, we would expect to see the inferred metallicity vary between the models where we freeze the soft protons and the models where we allow some freedom in fitting this background; no such variation is observed between these models. 

An incorrect neutral Hydrogren column would also lead to incorrect estimation of the continuum. The Galactic $N_H$ column
was estimated from \citet{Dickey1990}, but \citet{Kalberla2005} also finds a similar value, indicating that the Galactic column is not particularly uncertain at this location. We checked the Planck CO maps as well and see no molecular gas in the direction of NGC 1961. 

Finally, NGC 1961 contains significant amounts of neutral gas in addition to the hot X-ray gas. It is therefore plausible that interactions between these phases would produce charge exchange emission. Depending on the amount of this emission, and on the ratio of \ion{O}{vii} to \ion{O}{viii} emission, this effect might also change the ratio of the continuum to the Fe L complex. This sort of emission is known to be important in the starburst galaxy M82 (\citealt{Liu2011}, \citealt{Zhang2014}) and in a few other cases \citep{Li2013}, but a systematic study has yet to be performed so it is not yet possible to estimate its importance in NGC 1961. 

In summary, our inference of a low metallicity for the hot halo of NGC 1961 seems to be robust against basic systematic errors, but it is not definitive. We therefore include a factor of $Z/0.2 Z_{\odot}$ in subsequent figures to emphasize this uncertainty and show the effect of different choices on the proceeding conclusions. We also remind the reader that the Iron abundance in \citet{Anders1989} is about 40-50\% higher than in other abundance tables. Accounting for this difference makes our metallicity consistent with the $0.2-0.3 Z_{\odot}$ metallicity which is ubiquitously observed in the outskirts of galaxy clusters (\citealt{Bregman2010}, \citealt{Werner2013}).

\subsection{Two-temperature fits}
  
Here we explore the viability of two-temperature fits to the hot halo of NGC 1961, instead of relying on a single APEC model.  We focus on the 2' annuli in order to ensure sufficient photon statistics, and we use the models with frozen proton background and frozen XRB component. We initialize the two components at 0.3 and 0.8 keV and let the metallicity float. We therefore have three additional degrees of freedom (temperature, metallicity, and normalization of the second hot gas component) as compared to the one-T model, so for the second component to be statistically significant at $p=0.99$, we require it to improve the $\chi^2$ by at least 11.35.

\begin{table*}
\begin{minipage}{135mm}
\caption{2-Temperature spectral fits}
\begin{tabular}{clllllll}
\hline
region  & $T_1$ & $Z_1$ & $\int n_en_HdV$ &$T_2 $ & $Z_2$ & $\int n_en_HdV$ & $\Delta \chi^2$\\
 & (keV) & ($Z_{\odot}$) & ($10^{62}$ cm$^{-3}$) &  (keV) & ($Z_{\odot}$) & ($10^{62}$ cm$^{-3}$) &  \\
\hline
R012  & $0.75^{+0.09}_{-0.04}$ & $0.19^{+0.07}_{-0.06}$ & $71.5^{+20.4}_{-24.8}$& $0.25^{+0.10}_{-0.07}$& $1.54^{+0.85}_{-0.82}$ & $6.7^{+4.6}_{-2.5}$ &30.8 \\
R123 &  $0.74^{+0.19}_{-0.11}$ & $0.18^{+0.10}_{-0.08}$& $36.8^{+29.6}_{-16.0}$ & $0.28^{+0.07}_{-0.09}$ & $1.96^{+1.14}_{-1.64}$ & $5.3^{+21.3}_{-2.6}$ &33.3 \\
R234  & $0.35^{+0.12}_{-0.07}$& $0.25^{+0.43}_{-0.18}$ & $35.4^{+40.9}_{-22.8}$ & $1.00^{+1.00}_{-0.35}$ & $1.73^{+2.67}_{-1.18}$ & $1.8^{+1.5}_{-1.3}$ &24.3 \\
R345 &  $0.38^{+0.13}_{-0.05}$ & $0.27^{+0.34}_{-0.15}$ & $17.3^{+18.0}_{-9.2}$ &$0.02^{+0.01}_{-0.01}$& $2.30^{+2.39}_{-2.12}$ & $1.2^{+1.2}_{-0.01}\times10^5$ &  2.2  \\
R456&$0.35^{+0.17}_{-0.08}$ & $0.38^{+1.00}_{-0.32}$ & $7.5^{+24.1}_{-5.2}$& $0.02^{+0.01}_{-0.01}$& $0.01^{+0.01}_{-0.01}$ & $5.5^{+6.0}_{-2.2}\times10^6$ & 0.2\\
\hline
\end{tabular}
\\
\small{Two-temperature fits to the 2' regions. The final column shows the improvement in the $\chi^2$ goodness of fit parameter for the 2-T model relative to the 1-T model; a value of at least 11.35 corresponds to an improvement significant at more than 99\% confidence. The 2-T model is statistically favored in the inner three regions, which include the disk of the galaxy, but offers no significant improvement in the outer regions. }
\end{minipage}
\end{table*}

The inner three regions (which overlap with the disk of the galaxy) show significant improvement with a 2-T model, while the outer two regions (which only cover the hot halo) do not show much improvement with a 2-T model. This seems intuitively reasonable, and one can imagine that one component could describe the hot halo and the other could describe the hot ISM of the disk. In general, the first component has a low metalllicity and contains most of the mass; its temperature, metallicity, and density behave like the 1-T fits shown in the previous section. This component can be associated with the hot halo of the galaxy. Note that, while the emission measures in Table 3 are larger than the fiducial emission measures, this difference stems from the lower metallicites in the 2-T fit as compared to the fiducial model (which has $Z\equiv0.5 Z_{\odot}$). The second component is more poorly constrained, but generally has a higher metallicity and a lower emission measure. This component can be associated with the hot interstellar medium of the galaxy. The 2-T model therefore seems to broadly support our picture of the hot gas in and around NGC 1961. Unfortunately there are not sufficient photons to do a more detailed analysis of the ISM of this galaxy at present.

\section{Spatial Analysis}

In this section we examine the surface brightness profile of the X-ray emission around NGC 1961. We analyze the soft-band and hard-band images generated in Section 2, as well as model images of the various background components. The ESAS \verb"mos_back" and \verb"pn_back" routines generate model particle background images automatically, and we use the fit parameters from our fit to the background at large radii as inputs to the the ESAS \verb"soft_proton" and \verb"swcx" routines in order to generate models of the soft proton background and the SWCX backgrounds for each observation and each detector (these two backgrounds are only generated for the soft band, however). For each type of image, we use the \verb"merge_comp_xmm" routine to combine all 30 images together into a single merged image. We also use this routine to combine the 30 exposure maps together to generate a merged exposure map. We present the merged, exposure-corrected, soft-band image in Figure 5, which was created with the \verb"merge_comp_xmm" routine and has the QPB, soft proton, and SWCX backgrounds all subtracted. Note that {\bf this routine applies no weighting between the MOS1, MOS2, and PN detectors, but instead adds the images (in counts) directly}; this has the effect of weighting the PN image more than the MOS images due to the larger effective area of the PN detector.

\begin{figure*}
\begin{center}
{\includegraphics[width=10cm]{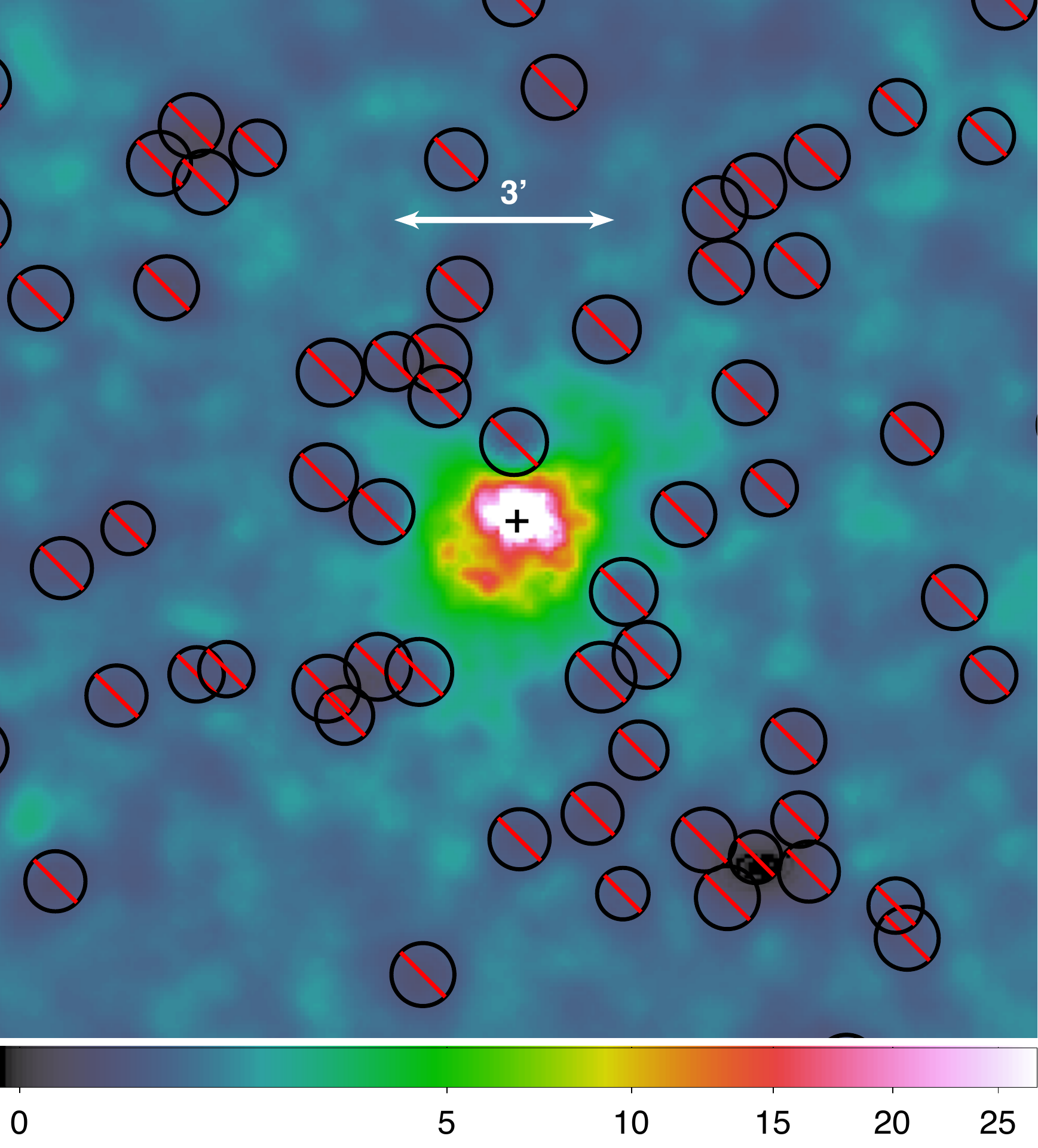}}
\end{center}
\caption{Adaptively smoothed merged XMM-EPIC image of the 0.4-1.25 keV emission from the region around NGC 1961. This image has been exposure-corrected and the estimated particle background, soft proton background, and SWCX background have been subtracted; point sources have also been masked. The CXB, Galactic halo, and Local Bubble have not been subtracted and these backgrounds produce the uniform background which fills the field. The white arrow indicates 3', or about 50 kpc at the distance of NGC 1961, and the black cross denotes the center of the optical emission from the galaxy. The hot halo is visible by eye to several arcminutes and can be studied at larger radii through surface brightness profiles. }
\end{figure*}

We divide each image by the merged exposure map and construct surface brightness profiles in Figure 6. The net soft-band signal flattens to a constant value at large radii; we use this mean value as the value for the X-ray background in this field over the soft band. The X-ray background (the combination of the CXB, Galactic halo, and Local Bubble) is the largest background component, followed by the particle background, the soft proton background, and finally the SWCX background. The sky background estimated from Figure 6 is consistent with the 3/4 keV background for the same region measured from ROSAT as well. Note that the particle background and the proton background appear to turn upwards at large radii; this is because these backgrounds are not focused by the telescope's mirrors and therefore do not suffer the same vignetting as the X-ray backgrounds. Dividing by the exposure map therefore overestimates these backgrounds at large radii. This should not be a major issue for our analysis, since it does not begin to matter until radii of 100 kpc or more, which is outside the range within which we can measure the hot halo.

In these profiles we also show, but do not subtract, the estimated contribution of X-ray binaries to the soft band. We estimate this contribution in two different ways. One method (the cyan line) uses the hard-band X-ray image and model backgrounds. We determine the surface brightness profile of the hard-band emission and multiply this profile by a factor of 1.67 (the scaling into the soft-band summed MOS1+MOS2+PN counts for a $\Gamma=1.56$ powerlaw with Galactic absorption) to generate the estimated XRB profile in the soft band. For the other method (magenta line), we use the K-band image of this galaxy from 2MASS \citep{Skrutskie2006} to estimate the K-band surface brightness profile, then convert to the soft band using the scaling relation for LMXBs from \citet{Boroson2011} (this conversion is also described in more detail in \citealt{Anderson2013}). These two methods agree reasonably well, both showing a subdominant but somewhat uncertain XRB contribution in the central region, which falls off quickly beyond about 1.5 arcmin (23 kpc) and becomes insignificant at larger radii.

\begin{figure*}
\begin{center}
{\includegraphics[width=16.5cm]{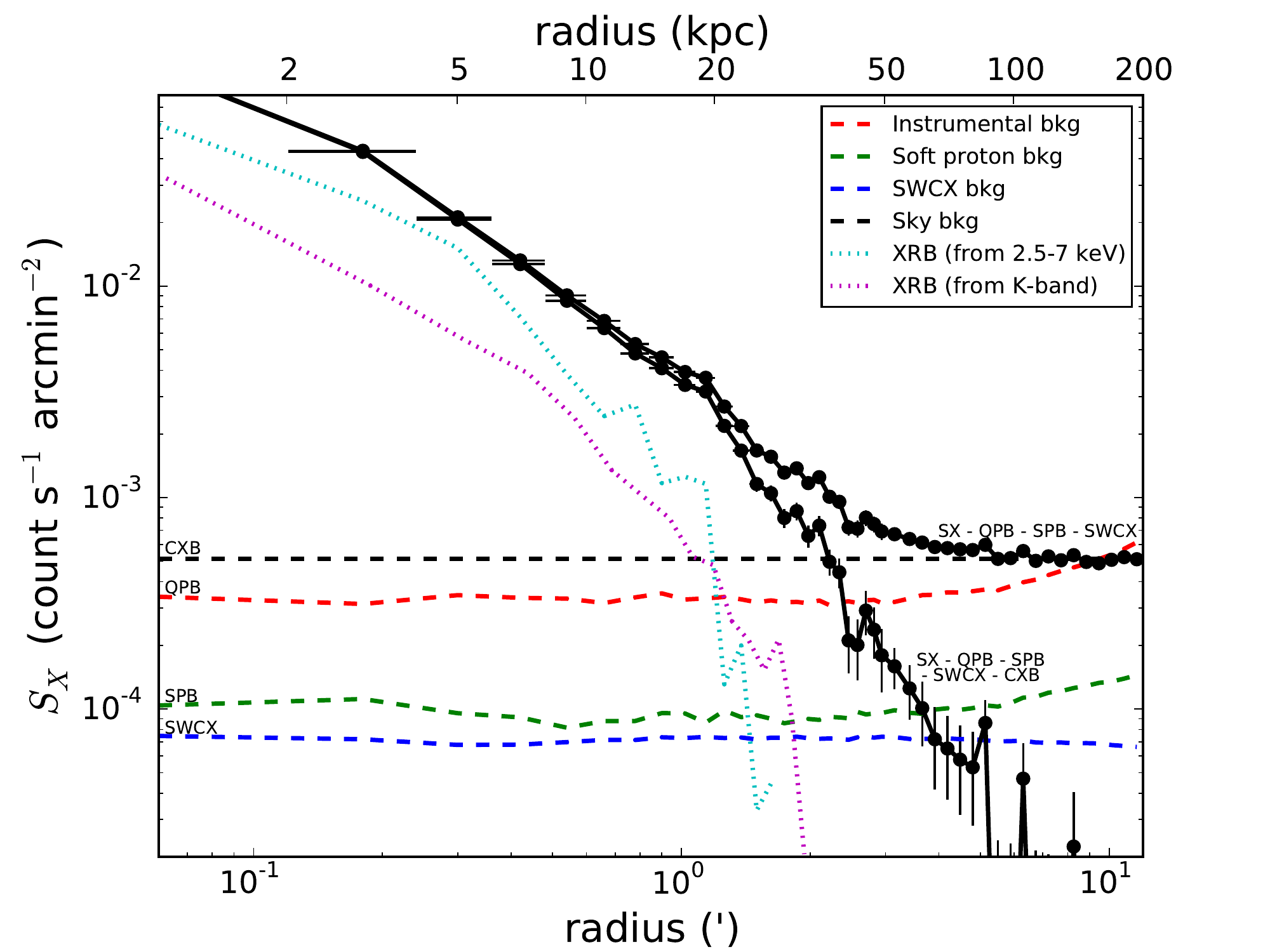}}
\end{center}
\caption{Background-subtracted surface brightness profile of the 0.4-1.25 keV emission around NGC 1961. The black points are the background-subtracted data, showing the remaining emission after subtracting: (red) particle background, (green) soft proton background, and (blue) Solar wind charge exchange background. The upper set of black points include the contribution of sky background, which is assumed to be flat across the field and fit with the dashed black line. The lower set of black points has the estimated sky background removed as well, leaving only the emission we attribute to NGC 1961. Finally, the dotted cyan and magenta lines show the estimated contribution of X-ray binaries in NGC 1961, using two different methods. These are not subtracted from the profile but are clearly subdominant in the soft band.}
\end{figure*}

We also verify that our results are consistent between the spectral and spatial analyses. This is important to check (Anderson and Bregman 2014), especially with the complexity of our background model. To check this, we plot the surface brightness profile of the hot gas as determined from each technique, in Figure 7. The black line shows the result of the spatial analysis, which is the remaining soft-band emission after subtracting the estimated CXB, quiescent particle background, soft proton background, Solar wind charge exchange background, and XRB emission as estimated from the hard-band image (the cyan line in Figure 6). We restrict this plot to the MOS instruments, since the PN instrument has a different effective area. The red points are the results of the spectral analysis (using the fiducial model), where we have converted the APEC component describing the hot halo (including Galactic and intrinsic absorption) into a soft-band count rate for the MOS detectors.

The agreement is excellent within the $\sim$ 42 kpc where the spectral fits are robust. In the outer spectral regions, where the significance of the hot gas component is less than 3$\sigma$, the spectral model predicts far lower surface brightness than the observed spatial profile. However, here the hot gas emission comprises less than about 10\% of the total soft-band signal, and it is not possible to say with certainty what the true emission looks like at such low surface brightnesses. We discard these outer regions from the joint analysis below. 

\begin{figure}
\begin{center}
{\includegraphics[width=8cm]{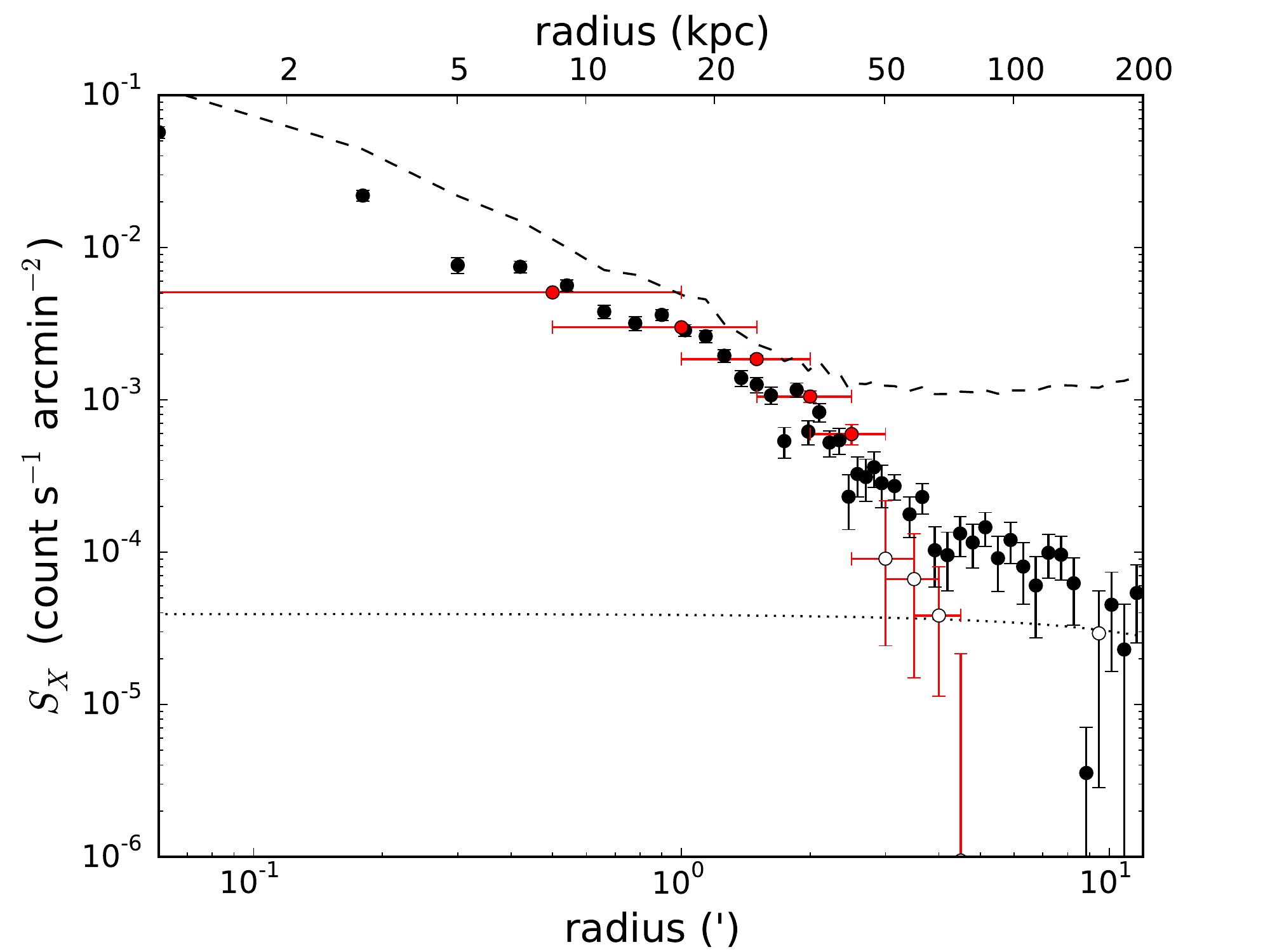}}
\end{center}
\caption{Spatial (black points) and spectral (red points) comparisons for the MOS soft-band surface brightness profile attributed to the hot gaseous halo. For comparison, we also show the soft-band surface brightness profile of the full image (i.e. instrumental + sky backgrounds; dashed line) and the surface brightness profile from a model hot halo with a uniform density of $200\rho_c \Omega_b$ (dotted line). Black points with open circles denote negative values which have been multiplied by $-1$ to allow plotting with a logarithmic y-axis. Red points with open circles are spectral fits where the significance of the hot gas component is below 3$\sigma$. The results from the spectral and spatial methods largely agree within $r \lapprox 42$ kpc. At larger radii, where the surface brightness of the hot gas is below about 10\% of the total signal, we cannot tell which is correct: the profile from the spatial method, the profile from the spectral method, or neither. }
\end{figure}

\section{Physical Properties of the Hot Halo}
Now that we have performed both a spectral analysis and a spatial analysis, and shown that they give consistent results, in this section we explore the physical properties of the hot halo. First, we deproject the spectral results in order to estimate the hot gas density profile. Conceptually, we follow a similar procedure to that of \citet{Churazov2003}, although these data have much lower S/N than their observations of the Perseus cluster so we make a few modifications to that procedure.

First, we generate a simple estimate for the emission measure of the hot halo at very large radii. To do this, we fit a power-law to the hot gas component of the observed surface brightness profile at large radii, finding a logarithmic slope of $3.5$ for the surface brightness as a function of radius. This corresponds to a slope of $2.25$ for the density as a function of radius, which is equivalent to $\beta$ of 0.75 in the standard $\beta$-model. This conversion between surface brightness profile and density profile assumes that the emissivity does not vary with radius. In our 0.4-1.25 keV band, the emissivity does not change with temperature by more than 10\% for temperatures between 0.3 keV and 1 keV. If the metallicity varies, the emissivity can also change, but for this galaxy the projected metallicity profile is consistent with being flat. We use a fiducial temperature of 0.4 keV and a fiducial metallicity of 0.2 $Z_{\odot}$ to convert from surface brightness into density. 

Next, we convert the results of the spectral fitting into estimates of the emission measure in each region. This is trivially derived from the normalization of the APEC component of the spectrum, using our assumed distance of 58.0 Mpc to NGC 1961. We rescale the emission measure in each region assuming the fiducial metallicity of 0.2$Z_{\odot}$. While we adopted a metallicity of 0.5$Z_{\odot}$ when performing our spectral fitting, the results of the spectral fits with floating metallicity suggest that a value of 0.2$Z_{\odot}$ is more reasonable for this galaxy. {\bf We therefore adopt a fiducial metallicity of 0.2$Z_{\odot}$ for the remainder of the analysis}. It is simple to scale our results into a different metallicity, however: at the level of accuracy we can measure for this galaxy, a good approximation is that the emission measure is inversely proportional to the value of the metallicity, and the electron density is proportional to the inverse square root of the metallicity. 

We treat the spectral fits to the 1' bins and the 2' bins separately, and since the regions overlap we also separate each set into two groups. We therefore have four independently determined profiles, based on regions R0-R2-R4, R1-R3, R012-R345, and R123-R456 respectively. As noted in section 4, we discard the R5-R8 regions since the normalization is so poorly constrained from the spectral fitting in these regions.

Finally, for each annulus, we subtract the expected emission measure from exterior annuli (EM$_{\text{ext}}$). We divide the remaining emission measure by the volume of the annulus, scaled by the fraction of the field of the annulus covered by the fiducial spectrum (using the area $A$ listed in Table 2). We also include a factor of 0.83 to convert from $n_H$ into $n_e$ assuming a standard Helium abundance. Putting this all together, and using a distance of 58.0 Mpc and angles in units of arcminutes, the expression for the average electron density is (see also \citealt{McLaughlin1999}, who derives a more general form of this equation):

\begin{equation}
n_e = 8.00\times10^{-2} \text{ cm}^{-3} \times \sqrt{ \frac{\int{n_e n_H} dV - \text{EM}_{\text{ext}}}{10^{66} \text{ cm}^{-3}} \frac{\theta_2^2 - \theta_1^2}{A\left(\theta_2^3 - \theta_1^3\right)}}
\end{equation}

These values are listed in the final column of Table 2, and displayed graphically in Figure 8.  The systematic errors seem to be larger than the 1$\sigma$ statistical uncertainties, but together they are still only at about the 10\% level. The uncertain metallicity  is by far the largest source of uncertainty. We also note that in the inner three regions, deviations from hydrostatic equilibrium may be expected due to the presence of the galactic disk. There is some evidence for this in the preference for 2-temperature fits to these spectra, but the underlying hot halo component remains dominant in these regions, and there is no evidence of the disk in the X-ray image (Figure 5).

\begin{figure*}
\begin{center}
\subfigure[1' annuli]{\includegraphics[width=8cm]{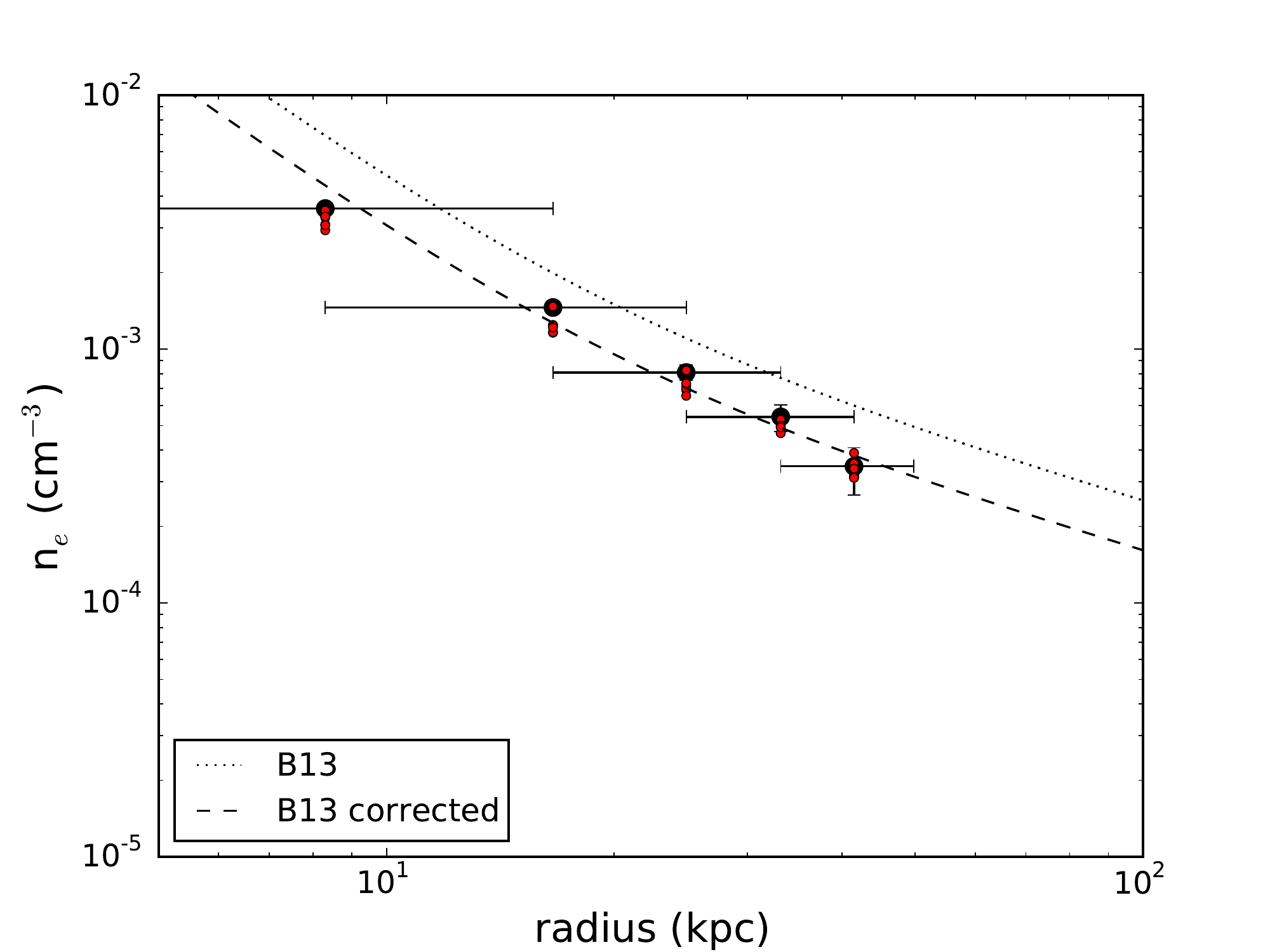}}
\subfigure[2' annuli]{\includegraphics[width=8cm]{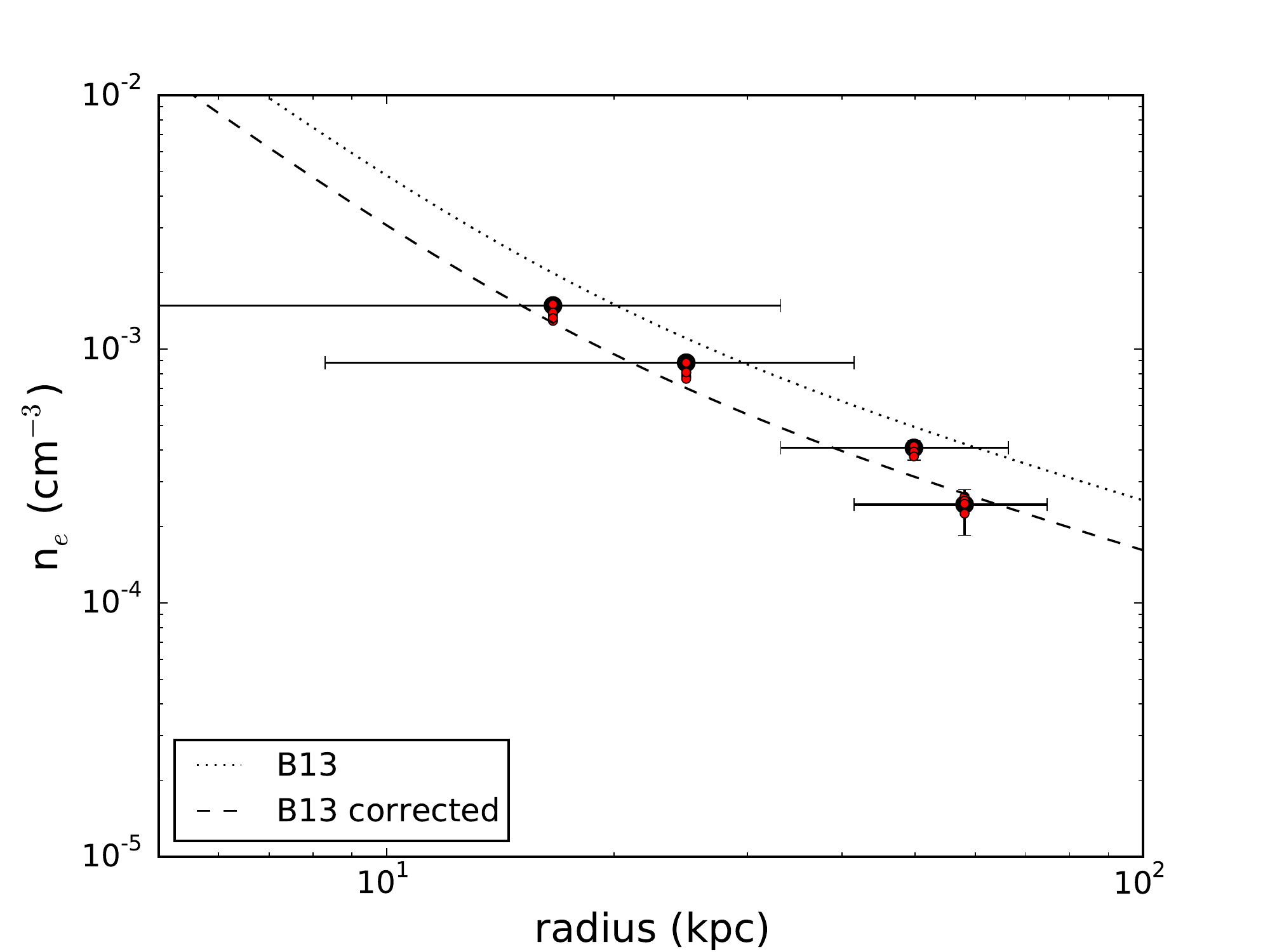}}
\end{center}
\caption{Deprojected electron density profiles of the hot halo of NGC 1961. Since the data points overlap, each plot shows two independent profiles, derived from non-overlapping points. The two independent profiles are perfectly consistent with one another. As in the other figures, the black points show the results for our fiducial model and the red points show the results for the other spectral models, and the error bars are 1$\sigma$ around the median  for the fiducial model. For comparison, the density profile for this galaxy measured by B13 is displayed as well, along with a ``corrected'' version of their density profile rescaled to match our fiducial metallicity of $0.2 Z_{\odot}$. The corrected profile matches our results well. }
\end{figure*}

In Figure 8 we also plot the best-fit hot halo electron density profile for this galaxy, as measured by B13. They used the modified $\beta$-model profile of \citet{Vikhlinin2006} to parameterize the surface brightness profile, and assumed a constant metallicity of $0.12 Z_{\odot}$ (relative to the abundance tables of \citealt{Grevesse1998}). We also compute a ``corrected'' version of their density profile, which is multiplied by a factor of 0.64 to account for the different metallicity and the different abundance table relative to our analysis. Overall the agreement is very good between our deprojected profile and their corrected best-fit profile. The behavior of the profile at larger radii is extremely important, however, and it is not clear whether their parameterization can be extended to larger radii. Improved observations are necessary in order to understand the behavior of the hot gaseous halo within a larger fraction of the virial radius. 

\subsection{Pressure and Mass Profiles}
We can estimate an electron pressure profile for the hot gas, which is the product of the electron density and the temperature. Unfortunately, we do not have a deprojected temperature profile, and the hot gas component of the spectra is subdominant beyond a projected radius of 2 arcminutes, so we cannot get robust results by subtracting scaled spectra from one another and fitting the remainders, as one can do for deprojection in the high S/N regime. However, our projected temperature profiles do not show significant gradients, and we can produce an approximate estimate for the reprojected pressure profile by multiplying the reprojected density profile by the projected temperature profile. We propagate the uncertainties on the temperature and the density into the total uncertainty, and neglect the additional uncertainty introduced by using a projected temperature profile instead of a deprojected temperature profile (this should be smaller than the statistical uncertainties on the temperature, however). The resulting profile is shown in Figure 9.

\begin{figure}
\begin{center}
{\includegraphics[width=8cm]{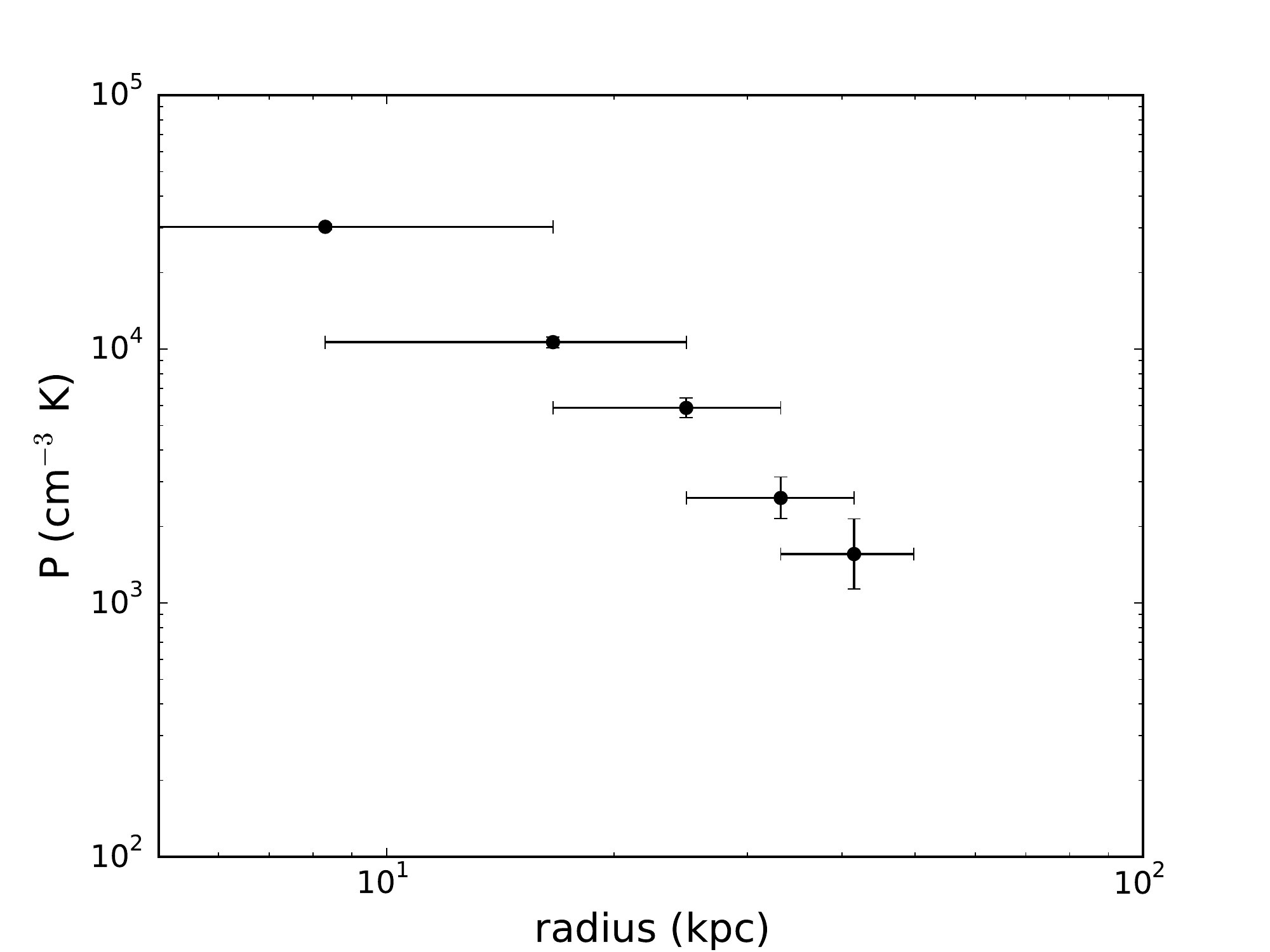}}
\end{center}
\caption{Approximate electron pressure for the hot halo of NGC 1961, based on the temperature profile and the deprojected electron density profile (see section 5.1). Errors are $1\sigma$.  }
\end{figure}

Now that we have deprojected electron density and electron pressure profiles, we can estimate the total mass profile of the galaxy. This is derived from the temperature, electron density, and the gradient of the total pressure at each radius. We assume $\mu=0.61$ and we assume the total pressure is 1.91 times the electron pressure. This factor is appropriate for Solar metallicity gas with the \citet{Anders1989} abundances, but does not vary by more than 1\% for other reasonable abundances. We neglect magnetic support or other forms of non-thermal pressure. We interpolate between the points in our pressure and density profiles in order to calculate the gradient, and extrapolate to larger and smaller radii based on the distance-weighted slope of the nearest data points (for more details, see Appendix B of \citealt{Churazov2008}). The effective circular velocity corresponding to this derived mass profile is shown in Figure 10. 

For comparison, we also plot a number of other constraints on the circular velocity profile of this galaxy. \citet{Haan2008} measured the HI circular velocity out to 43 kpc, with independent measurements for the receding and approaching sides of the disk. They measure an inclination angle of $i=42.^{\circ}6\pm4^{\circ}.0$ (close to the $47^{\circ}$ listed in HyperLeda) which they use to correct their measurements. The approaching side of the disk decreases towards zero velocity near the center, but the receding side of the disk shows a roughly flat curve towards the center. At smaller radii, CO 1-0 measurements are also available from \citet{Combes2009}, which are more consistent with the behavior of the receding side of the disk when corrected using $i=42.^{\circ}6$.

\citet{Combes2009} make an argument for a higher inclination angle, however, using $i=65^{\circ}$ as a rough value which we adopt here for comparison. This value would reduce the inclination correction and the inferred total mass for the galaxy. Applying this inclination correction seems to introduce some tension between the cold gas measurements and our X-ray measurements, however. 

We consider a third mass estimate using the K-band image to derive the stellar contribution to the potential. This yields a lower limit to the total circular velocity. We extract the stellar profile from the K-band image using ellipses for both $i=42.^{\circ}6$ and $i=65^{\circ}$, but the results are nearly identical in both cases so we only plot the former for simplicity.

Using the assumed M/L ratio of 0.6 from Section 1, which is roughly the expectation based on \citet{Bell2001} and \citet{Chabrier2003} for this galaxy, we see that the stars contribute the majority of the mass in the central region of the galaxy, and are also in tension with the gas-dynamical measurements corrected assuming $i=65^{\circ}$. If we instead consider the \citet{Salpeter1955} M/L ratio, the tension increases further. The \citet{Haan2008} inclination angle gives results which seem consistent among gas-dynamical measurements, X-ray measurements, and the stars. In the regimes where each curve is reliable, the agreement between the rotation curves is good, and points to a roughly flat rotation curve within at least 42 kpc. 

\begin{figure}
\begin{center}
{\includegraphics[width=8cm]{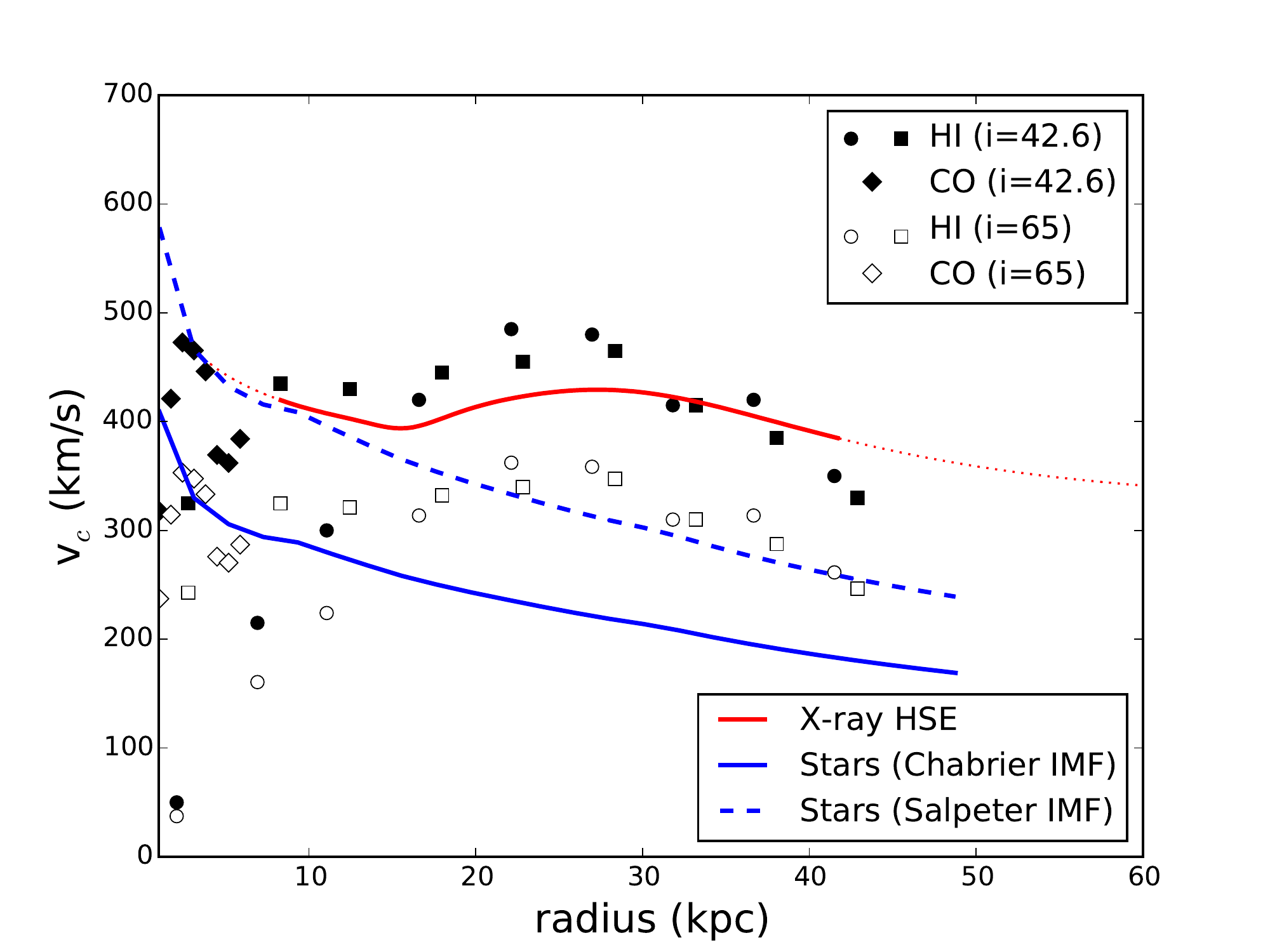}}
\end{center}
\caption{Rotation curves for NGC 1961. Data points are inclination corrected assuming $i=42.^{\circ}6$ or $i=65^{\circ}$, as indicated. HI data are form \citet{Haan2008} for the approaching (circles) and receding (squares) sides of the disk. CO 1-0 data are from \citet{Combes2009}. The red line is our estimate of the effective circular velocity based on our X-ray data, assuming hydrostatic equilibrium. The line is shaded thick over the region where the rotation curve is constrained by data, and dotted where the curve is extrapolated. The blue lines show the approximate contribution of the stars in NGC 1961, based on the K-band image assuming either a Chabrier (solid line) or Salpeter (dashed line) IMF. In general the $i=42.^{\circ}6$ model seems to be preferred by the X-ray data, and the rotation curve seems to be roughly flat within at least 42 kpc.}
\end{figure}

Finally, in Figure 11 we plot the enclosed stellar mass, hot gas mass, and total mass for NGC 1961. We use the aforementioned M/L ratio of 0.6 for the stars, so that the total stellar mass is about $3\times10^{11} M_{\odot}$. The rotation curves only extend to about 42 kpc, so we extrapolate the total mass profile to larger radii. We assume that the dark matter approaches an NFW profile at larger radii \citep{Navarro1997}, such that the virial mass of this galaxy is $1.3\times10^{13} M_{\odot}$, extending to a virial radius of about 490 kpc. This agrees well with the estimate of B13, who estimated a virial radius of 470 kpc for this galaxy based on comparison with cosmological simulations. Still, it is a very crude estimate of the total mass, and we will therefore not use the extrapolated mass for any precise calculations. 

Within 10 kpc, the stellar component seems to be dominant, but at larger radii the system quickly becomes dominated by dark matter. Within 90 kpc, the hot halo contains less than a tenth of the mass in the stars. At this radius, the baryon fraction (mass in stars + neutral Hydrogen + hot halo) is close to the Cosmic fraction. Extrapolating to larger radii, the hot halo may become comparable to the mass of the stars, but the sum of these components is less than a third of the expected baryon content for the system. This is an illustration of the problem of missing baryons from galaxies (see section 6.1). 

\begin{figure}
\begin{center}
{\includegraphics[width=8cm]{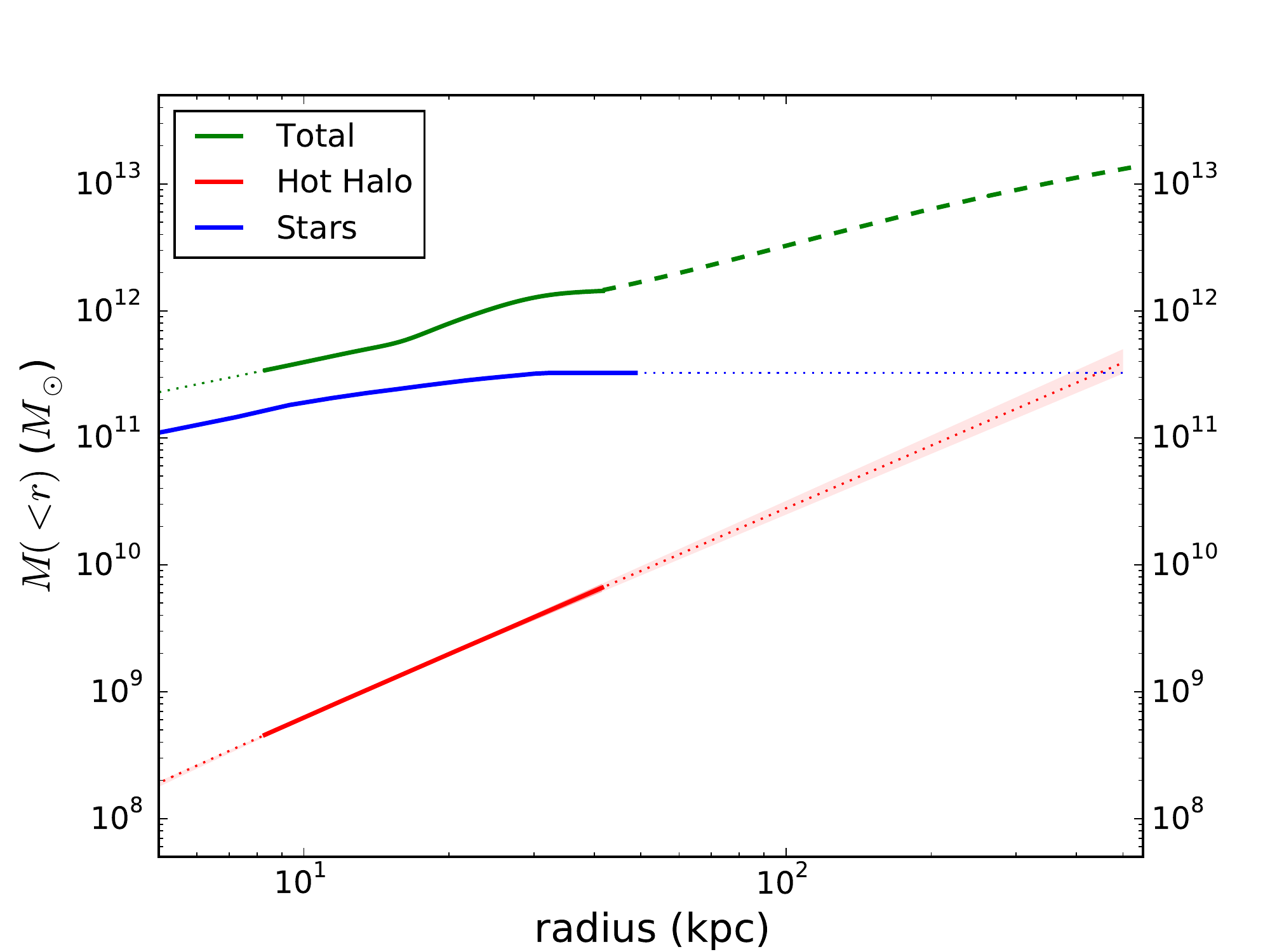}}
\end{center}
\caption{Enclosed mass profiles for NGC 1961. The green line shows the total mass, as inferred from the X-ray observations using the assumption of hydrostatic equilibrium (section 5.1). The dashed extension at large radii is an NFW profile fit to the observed portion of the mass profile, as described in the text. The blue line is the stellar mass, as inferred from the 2MASS K-band image of this galaxy using a M/L ratio of 0.6. The red line is the mass of the hot gaseous halo, including $1\sigma$ uncertainties as the shaded red region. Each line is styled in boldface over the regime where the mass is measured, and dotted where the profile is extrapolated.  }
\end{figure}

Our measured electron pressure profile is also useful for predicting the thermal SZ signal from this galaxy. The thermal SZ effect is proportional to the volume integral of the electron pressure, which we express using the Compton $y$-parameter. The Compton $y$ parameter for our pressure profile, integrated over a sphere with radius 42 kpc, is $2\times10^{-6}$ arcmin$^2$. The typical radius for comparison with other studies is $R_{500}$, which for our profile is $\sim$325 kpc ($19.6$') for NCG 1961 (and the corresponding $M_{500}$ is $1\times10^{13} M_{\odot}$). If we extrapolate our pressure profile out to this radius, the integrated $y$ parameter increases to $1\times10^{-5}$ arcmin$^2$. These values are below the sensitivity limits of Planck SZ catalogs for an object of this angular size (\citealt{Planck2015b}, \citealt{Khatri2015}) and the projected location of NGC 1961 is fairly close to the Galactic disk, which is a more difficult location in the sky to isolate the SZ signal. 

We can also compare this value to the stacking measurements of \citet{Planck2013}. For galaxies with log $M*/M_{\odot} = 11.5$, \citet{Planck2013} estimate $\tilde{Y}_{500} \approx 4\times10^{-5}$ arcmin$^2$, where the tilde indicates that the value has been normalized to an angular distance of 500 Mpc. Using the distance of 58 Mpc appropriate for NGC 1961 yields $Y_{500} \approx 3\times10^{-3}$ arcmin$^2$. This is about 300 times larger than our estimated $Y_{500}$ for this galaxy. Based on the Planck selection function data, for a galaxy with the angular size of NGC 1961, the Planck map should be sensitive to such a signal at 5$\sigma$ or higher with around 90\% completeness, so the non-detection of this galaxy suggests it does not have the mean thermal SZ signal for a galaxy of its stellar mass. On the other hand, based on the halo mass of NGC 1961, \citet{Planck2013} predicts $Y_{500}$ of just $1\times10^{-4}$ arcmin$^2$, which is undetectable in the Planck map. While this is only one galaxy, so it is difficult to draw strong conclusions, the implication is that halo mass is a better predictor of the SZ signal than the stellar mass, a point to which we return in section 6.2. However, the predicted $Y_{500}$ for a galaxy with the halo mass of NCG 1961 from \citet{Planck2013} is still an order of magnitude above our predicted $Y_{500}$ value (which admittedly involves an extrapolation of nearly an order of magnitude in radius beyond our outermost spectral bin). More work needs to be done in order to understand this discrepancy.

\subsection{Entropy Profile and HI Emission}

We can also construct the entropy profile for the hot halo of NGC 1961. This is given by $ K = T n_e^{-2/3}$, and propagates uncertainties on both $T$ and $n_E$ just like the pressure profile (Figure 9). The entropy profile is shown in Figure 12.

\begin{figure}
\begin{center}
{\includegraphics[width=8cm]{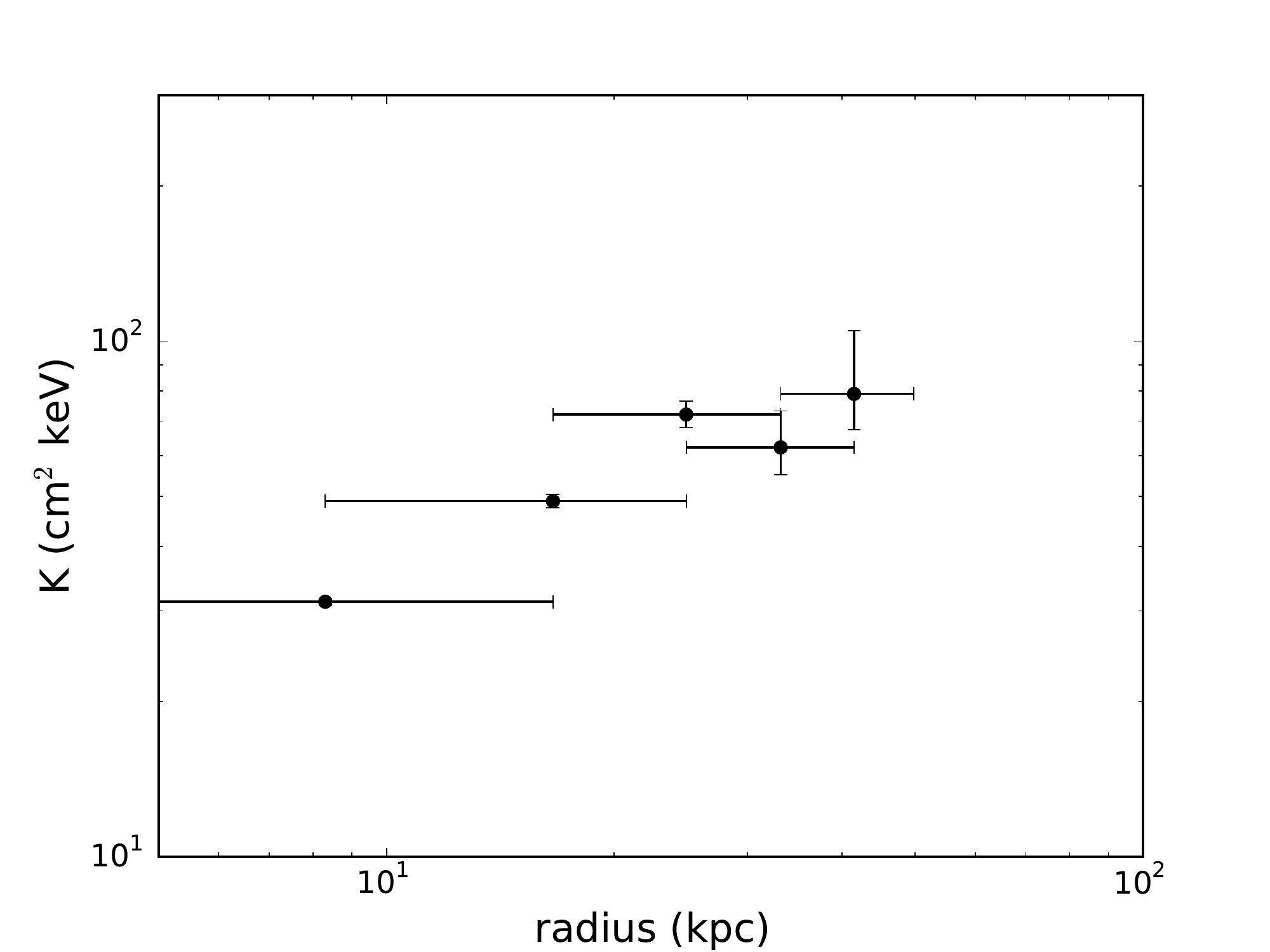}}
\end{center}
\caption{Approximate entropy profile for the hot halo of NGC 1961, based on the temperature profile and the deprojected electron density profile (see section 5.2). The entropy rises with radius, as expected for a dynamically stable halo. The central entropy is very close to the transition between cool-core and non-cool-core systems.  }
\end{figure}

Within the uncertainties, the entropy profile is positive in agreement with the Schwarzschild stability criterion. The central value is about 30 keV cm$^{2}$, which is the transition value noted by \citet{Cavagnolo2008} for brightest cluster galaxies (BCGs). BCGs below this value typically have bright H$\alpha$ emission and star formation, and generally correspond to cool-core systems \citep{Cavagnolo2009}, while higher entropy values indicate that the gas is able to stratify before cooling instabilities grow (\citealt{McCourt2012}, \citealt{Sharma2012}, \citealt{Gaspari2012}, \citealt{Voit2015}). The low entropy in the core of NGC 1961 therefore leads one to expect similar behavior for this galaxy. Indeed, 21 cm HI observations of this galaxy \citep{Haan2008} show widespread, patchy HI emission around this galaxy, extending tens of kpc to the NW. This extended HI emission has a low velocity dispersion (of order 20 km/s) and appears to be co-rotating with, and accreting onto, the disk (see Figure 6 of \citealt{Haan2008}). This extended HI appears to contain about 10\% of the total neutral Hydrogen content of the system, or about $5\times10^9 M_{\odot}$ of neutral gas. In order to check if this neutral gas can be explained by condensation from the hot halo, we compute the cooling time and the dynamical time as a function of radius. The cooling time is given by:

\begin{equation}
\tau = \frac{3/2 n k\text{T}}{\Lambda n_e (n - n_e)} \approx \frac{3/2 k\text{T} \times 1.91}{\Lambda n_e \times 0.91}
\end{equation}

\noindent where $n$ is the total particle density, and we assume $n = 1.91 n_e$ as in section 5.1. The dynamical time is given by

\begin{equation}
\tau = \sqrt{\frac{3\pi}{16 G {\bar{\rho}}}} =  \sqrt{\frac{\pi^2 R^3}{4 G M(<R)}}
\end{equation}

At every radius where we have data, the cooling time is about two orders of magnitude higher than the dynamical time, which is an order of magnitude away from the threshold value predicted from simulations for cooling instabilities to become important. At 40 kpc, for example, the cooling time is 13 Gyr and the dynamical time is 160 Myr. Therefore it seems unlikely that condensation from the hot halo is the source of most of the neutral Hydrogen seen around NGC 1961.

Another possibility, proposed by \citet{Shostak1982} is that the HI is the result of ram pressure stripping. NGC 1961 lies several hundred kpc to the NW of a poor group of galaxies, and the HI appears extended in the direction away from the center of this group, so it appeared plausible that NGC 1961 could be falling into this group and losing its ISM through interactions with a hot intragroup medium. However, the X-ray halo of NGC 1961 shows no signs of such interactions, and there is also no X-ray emission associated with a larger intragroup medium in Figure 5. A ROSAT analysis by \citet{Pence1997} also found no sign of an intragroup medium emitting in the X-ray band. 

We think the extended HI feature is more likely to be a sign of accretion from the intergalactic medium. This would explain its filamentary appearance, and it seems logical that an unusually massive spiral galaxy would require an unusually large supply of intergalactic gas. Moreover, the velocity field of the HI stream appears identical to the velocity field of the HI disk (see \citealt{Haan2008} Figure 6), which was also pointed out by \citet{Shostak1982}. Alignment of the angular momentum of the accreting gas and the galactic disk is the natural expectation for accretion from the IGM \citep{Mo1998}. Such alignment is less likely for a minor merger, so we think accretion is a better explanation than a merger for this galaxy, but a merger or accretion of a satellite galaxy is also a possibility. 

It is interesting that, although the entropy profile looks like a classic cool-core system, the temperature of the hot gas actually increases in the center by factor of 50\%. While cool cores are typical in galaxy groups and clusters, negative gradients are common for galaxies (\citealt{Humphrey2006}, \citealt{Diehl2008}), although X-ray faint ellipticals can show flat or positive gradients as well \citep{Fukazawa2006}. NGC 1961 is an isolated galaxy with the mass of a galaxy group, but it seems that the halo mass is not the determinant of the shape of the temperature profile for this system. We speculate that feedback from the galaxy itself (either stellar feedback or a previous episode of AGN feedback) may be responsible for heating the gas within the central few tens of kpc in this system.

\section{Discussion}

\subsection{Missing Baryons from NGC 1961}

In this section we return to one of the questions discussed in the introduction: where are the missing baryons from NGC 1961? The deep observations in this study allow us to constrain the density profile of the hot halo out to about 42 kpc, which is about 9\% of the virial radius, and it is notable that we see no evidence for any ``flattening'' in the density profile over this range.

It is still difficult to rule out a putative flat component at larger radii, however. As Figure 7 shows, the spectral and spatial techniques begin to disagree at radii larger than about 50 kpc, where the hot gas surface brightness is below about 10\% of the total soft-band surface brightness. We are unlikely to improve upon this limit (see, e.g. \citealt{Vazza2011}, \citealt{Reiprich2013}) with the current generation of instrumentation. In Figure 7 we also show the predicted surface brightness from a uniform density cloud of gas with kT = 0.5 keV, $Z = 0.2 Z_{\odot}$, and $n = 200 \rho_c \Omega_b$ extending out to the virial radius. This is the flattest possible density profile that contains the entire baryonic content of the system. Unfortunately this limiting case falls below our surface brightness sensitivity, and so we cannot confirm or rule out the possibility of some sort of extremely flat halo becoming important at larger radii. 

It is also important to emphasize the uncertainties in this discussion. We can detect the hot halo to nearly 20\% of the virial radius from our surface brightness analysis, and measure the density to about 10\% of the virial radius with good accuracy, but the vast majority of the virial volume is unconstrained and depends on power-law extrapolations. Moreover, the total mass of NGC 1961 is also not well-constrained, since it also depends on similar extrapolations to the observed rotation curves. Finally, since we find the stars have a similar mass as the hot halo for this galaxy, the M/L ratio of the stars is also somewhat important. Our estimates of each of these quantities are unlikely to be wrong by a factor of 2 or more, so the problem of missing baryons remains for NGC 1961, but the exact baryon fraction should be considered extremely uncertain.

Our measurements of the pressure in the hot halo can also be useful for understanding the baryon budget of NGC 1961. Observations with the COS instrument on HST show ultraviolet absorption around L* galaxies from many different elements, generally with a high covering fraction (\citealt{Tumlinson2011}, \citealt{Stocke2013}, \citealt{Werk2014}). Photoionization modeling suggests that this absorption comes from ionized clouds, with characteristic temperatures around $10^4$ K, densities of order $10^{-4}-10^{-3}$ cm$^{-3}$, and a volume filling fraction around 10\% \citep{Werk2014}. Such clouds cannot exist in pressure equilibrium with the hot halo around NGC 1961, but at larger radii (several hundred kpc) if the hot halo pressure continues to decline these clouds would be in equilibrium. If they have densities at the high end of their estimates and a 10\% volume filling fraction beyond 150 kpc, such clouds could contain essentially all of the missing baryons from the galaxy.  However, such clouds are unlikely to condense from the hot halo, given the high ratio of the cooling time to the dynamical time. If they originate from the galaxy, they would therefore have to survive the passage through the inner 100 kpc or more of hot gas.

\subsection{Comparison with Other Galaxies}

We measure the X-ray luminosity of the hot halo for this galaxy from the surface brightness profile (Figure 6). We convert from 0.4-1.25 keV counts into 0.5-2.0 keV flux assuming an APEC model with a fiducial temperature of 0.5 keV and a metallicity of 0.2 $Z_{\odot}$, although for reasonable values of $Z$ and $T$ the conversion factor is constant to within 5\%. The hot gas luminosity within 50 projected kpc is $8.9\pm1.2\times10^{40}$ erg s$^{-1}$. Integrating out further yields a total hot gas luminosity of $1.5\pm0.4\times10^{41}$ erg s$^{-1}$. This is slightly higher, but given the much larger uncertainties the difference is not very significant.

\begin{figure}
\begin{center}
{\includegraphics[width=8cm]{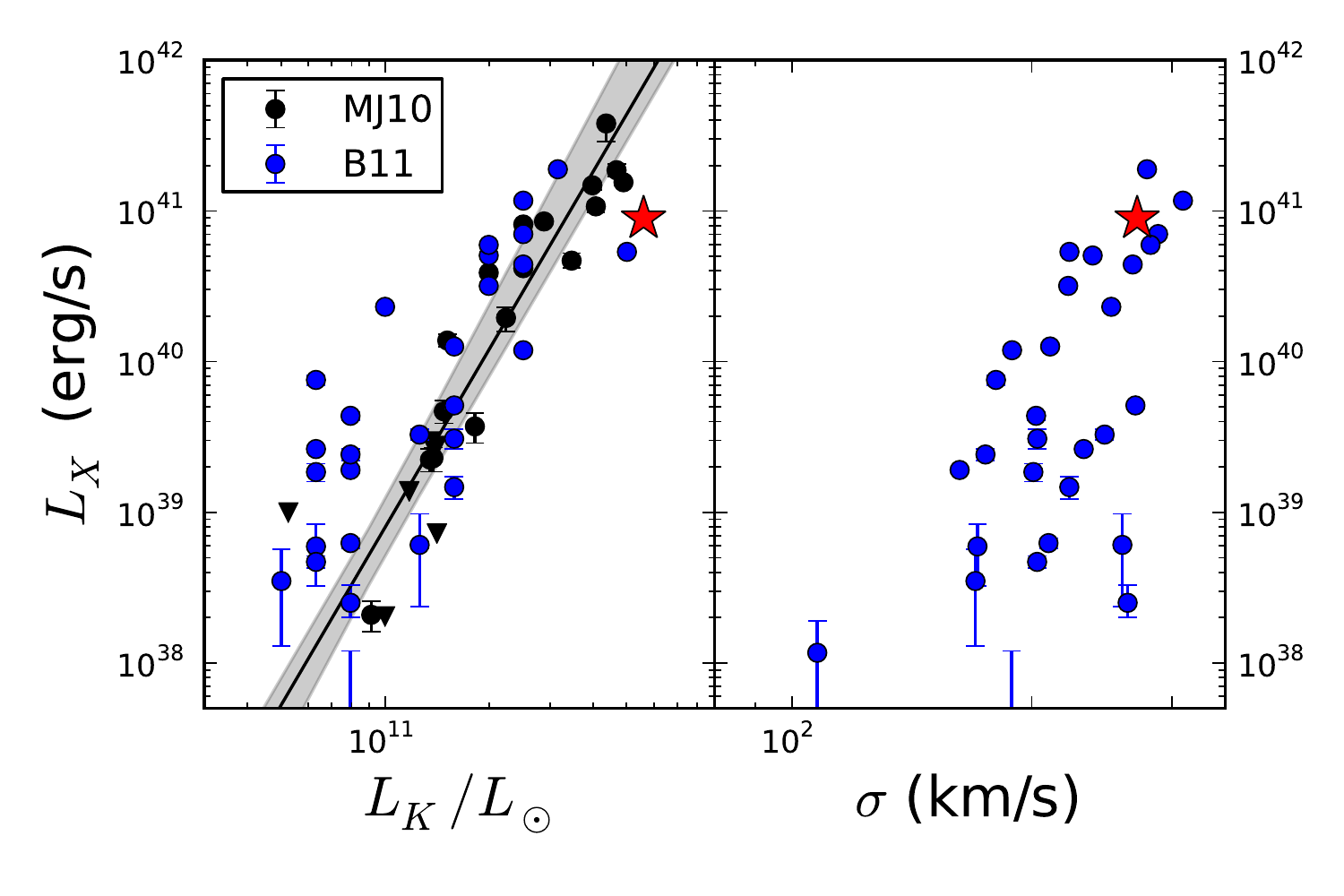}}
\end{center}
\caption{0.5-2.0 keV luminosity of the hot gas component of nearby galaxies, plotted against (left) K-band luminosity and (right) 1D stellar velocity dispersion.  Blue and black points correspond to field elliptical galaxies taken from the literature (\citealt{Mulchaey2010}, \citealt{Boroson2011}), and the black shaded region is the best-fit relation with 1$\sigma$ uncertainties from \citet{Mulchaey2010}. The red star is NGC 1961, using measurements from this work. We have converted the peak circular velocity from the HI profile into an effective velocity dispersion by dividing by $\sqrt{3}$. }
\end{figure}

In Figure 13 we compare the 0.5-2.0 keV luminosity of NGC 1961 (within 50 kpc) to various scaling relations for early-type galaxies. The left panel shows this galaxy on a plot of $L_X$ versus $L_K$, comparing against isolated galaxies from \citet{Mulchaey2010} and from \citet{Boroson2011}. NGC 1961 lies below their best-fit relation by about an order of magnitude. The scatter in these relations is also nearly an order of magnitude, so we caution against drawing strong conclusions from a single object, but this is consistent with the general observation that the X-ray halos of spiral galaxies appear less luminous than elliptical galaxies. 

NGC 1961 is much more consistent with relations between $L_X$ and the total matter content. In the right panel of Figure 13 we show the comparison between $L_X$ and the velocity dispersion of early-type galaxies from \citet{Boroson2011}. Since NGC 1961 is rotationally supported, we place this galaxy in the Figure by dividing the peak inclination-corrected circular velocity from the HI gas (Figure 10) by $\sqrt{3}$. In this case, NGC 1961 falls squarely within the relation for early-type galaxies.

We can show this in more detail with a detailed comparison between the properties of NGC 1961 and comparable elliptical galaxies (Figure 14). For this comparison we consider the galaxies NGC 57, NGC 4455, and NGC 7796, studied by \citet{OSullivan2004} and \citet{OSullivan2007}. They examine IC 1531 as well, but this galaxy has an AGN with a radio jet, and they cannot constrain the hot halo properties very precisely. These galaxies were selected as massive ellipticals which are not in any known group or cluster and which have no neighbors within 2 B-magnitudes of their own magnitude \citep{OSullivan2004}. \citet{OSullivan2007} note that the well-known elliptical NGC 720 falls near the border of their isolation criteria, but seems to have a higher density of faint galaxies around it than the three galaxies in their sample. As far as we can tell, NGC 1961 fulfills their isolation criteria as well, although it is less surprising for spiral galaxies to be isolated as compared to ellipticals.

\begin{figure}
\begin{center}
{\includegraphics[width=8.5cm]{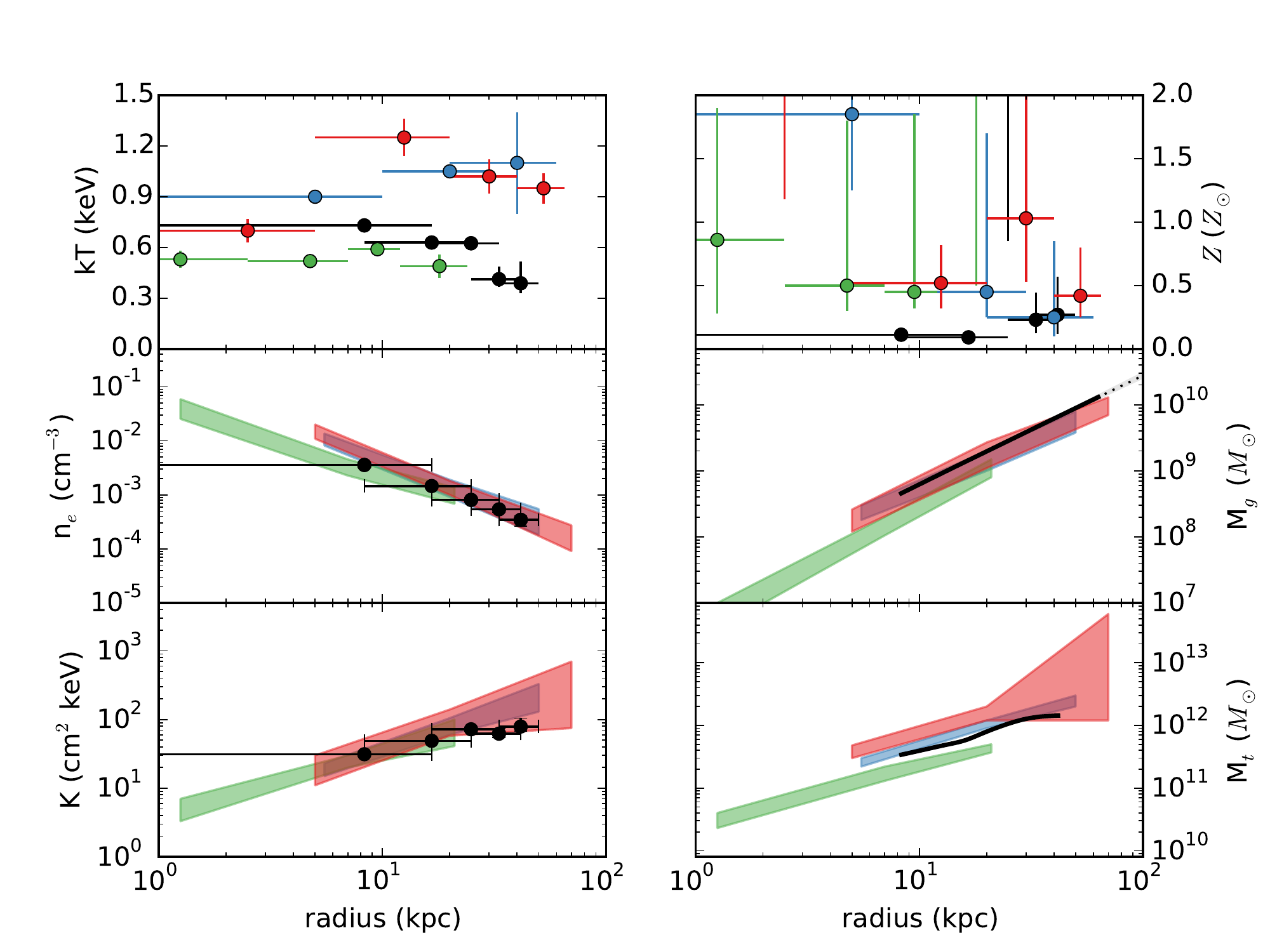}}
\end{center}
\caption{Temperature, abundance, density, enclosed gas mass, entropy, and enclosed total mass profiles for  NGC 7796 (green), (blue), and NGC 4455 (red), including 1$\sigma$ error regions, from O'Sullivan et al. (2004) and O'Sullivan et al. (2007). Values for NGC 1961 are shown in black for comparison.  }
\end{figure}

Within the uncertainties, the mass profiles and the entropy profiles for these four galaxies are all roughly consistent with one another. NGC 7796 has a slightly lower inferred total mass, so O'Sulivan et al. (2007) speculate that the hot gas around this galaxy might not be in hydrostatic equilibrium. There is a wider range of temperatures and abundances between these galaxies; NGC 1961 lies in the middle of the range of temperatures but has a lower abundance than the elliptical galaxies.  

An interpretation of the discrepancy with the $L_X-L_K$ relation and the agreement with the $L_X-\sigma$ relation is that the halo mass governs the X-ray properties of the system, instead of the stellar mass. As discussed in the Introduction, massive blue galaxies generally seem to lie in less massive halos than their red counterparts. As an example of this, from \citet{Anderson2015} we find that the effective halo mass of an average galaxy with the stellar mass of NGC 1961 is around $6\times10^{13} M_{\odot}$ - much higher than the halo mass we measure for NGC 1961. The lower halo mass offers a natural explanation for why NGC 1961 appears to scatter so low on the $L_X-L_K$ relation when compared to elliptical galaxies in much more massive halos.

\section{Conclusions}

In this paper we have analyzed the deepest X-ray observation to date of a hot gaseous halo around a spiral galaxy. NGC 1961 has already been studied with both Chandra and XMM-Newton, but our deep observations allow us to perform a much more precise and robust analysis. Using a spatial analysis similar to \citet{Anderson2011} and \citet{Bogdan2013}, we detect extended soft X-ray emission to very large radii (at least 80 kpc). No breaks or bright features are evident in the surface brightness profile or the soft X-ray image of the galaxy. 

We also perform a spectral analysis on the XMM-Newton data. This allows us to measure the physical properties of the hot gas out to 42 kpc. We explore eight different sets of spectral models, and the scatter between results from these different models allows us to estimate the systematic uncertainty from the choice of models. In general this is comparable to the statistical uncertainties. The spectral and spatial methods give fully consistent results as well, further underscoring the robustness of our results. 

The hot halo has a negative temperature gradient (declining with radius), as is typical for elliptical galaxies of similar mass. The metallcity of the halo is consistent with being flat at a value around 0.2$Z_{\odot}$. Such a low metallicity is not typical for elliptical galaxies of similar mass, which usually have abundances around Solar, although X-ray fainter ellipticals can have sub-Solar abundances as well. The higher abundances for ellipticals are thought to stem from the mixing of the hot interstellar medium and the hot halo, whereas NGC 1961 can lock up the majority of its metals in the cooler phases of its ISM. In support of this picture, we find a statistical preference for a two-temperature medium within the disk of this galaxy, with the dominant component having the high temperature and low abundance characteristic of the hot gaseous halo and a less massive component having a lower temperature and a higher abundance typical of the hot interstellar medium.

We also measure deprojected density profiles for the hot halo out to 42 kpc. The results are consistent with the parametric fit obtained by B13, when rescaled to our fiducial metallicity of 0.2$Z_{\odot}$, as well as with the much more uncertain profile obtained by \citet{Anderson2011}. No evidence of a break or a flattening in the profile is seen. Extrapolating the profile to the virial radius, we estimate a hot halo mass which is comparable to the stellar mass of the galaxy. 

From the density profile and the temperature profile, we compute a pressure profile for the hot halo and then we infer a total mass profile under the assumption of hydrostatic equilibrium for the hot gas. This mass profile is roughly consistent with measurements from the circular velocity profile observed with 21 cm emission. Based on the extrapolated hot gas mass and total mass, the problem of missing baryons from NGC 1961 seems to persist, with an estimated baryon fraction (hot gas + stars + neutral Hydrogen) around 30\% of the Cosmic fraction. 

It is conceivable that photoionized gas at intermediate temperatures could also exist around this galaxy, as observations by the COS instrument on HST suggest for many smaller galaxies. This could supply the missing baryons, but there are a few complications. First, such clouds would be far out of pressure equilibrium with the hot halo out to several hundred kpc. Second, the hot halo appears to be stable against global cooling instabilities over the range we can measure, making it difficult to condense such clouds out of the hot halo. These are not insurmountable difficulties, but they do complicate the modeling of the circumgalactic medium.

Finally, we show that the X-ray luminosity from NGC 1961 is about an order of magnitude fainter than one would expect from the $L_X$-$L_K$ relation for elliptical galaxies. On the other hand, this galaxy seems to have exactly the right $L_X$ for its dark matter mass, suggesting the dark matter halo governs the hot halo more fundamentally than the galaxy at the center. The non-detection of this galaxy in the Planck SZ map also supports this conclusion, as an elliptical galaxy with the same stellar mass and distance is very likely to have been detected.

A number of open issues remain. It is critically important to push measurements out to larger radii, which will require deeper X-ray observations, possibly combined with an SZ and/or absorption-line analysis (the latter two are more sensitive at larger radii than X-ray emission, which depends on the square of the gas density). It is also important to increase the sample size of hot gaseous halos, in order to understand better how the properties of the gaseous halo connect with the properties of the galaxy and of the dark matter halo.

\section{Acknowlegdements}
The authors would like to acknowledge helpful discussions with Mark Burke, Massimo Gaspari, Rishi Khatri, Alex Kolodzig, and Stefano Mineo related to this work, and to thank the referee for a thoughtful report that improved the paper. Based on observations obtained with XMM-Newton, an ESA science mission with instruments and contributions directly funded by ESA Member States and NASA.  This research has made use of the NASA/IPAC Extragalactic Database (NED) which is operated by the Jet Propulsion Laboratory, California Institute of Technology, under contract with the National Aeronautics and Space Administration. This research has made use of NASA's Astrophysics Data System. This research has made use of data and/or software provided by the High Energy Astrophysics Science Archive Research Center (HEASARC), which is a service of the Astrophysics Science Division at NASA/GSFC and the High Energy Astrophysics Division of the Smithsonian Astrophysical Observatory.

\end{document}